\documentclass[pra, twocolumn,amsmath, amssymb, notitlepage, longbibliography, showpacs, superscriptaddress]{revtex4-2}

\usepackage{graphicx}% Include figure files
\usepackage{dcolumn}% Align table columns on decimal point
\usepackage{bm}% bold math
\usepackage{float}
\usepackage[position=top,caption = false]{subfig}
\usepackage{textcomp} %\textmu
\usepackage{dsfont}
\usepackage[english]{babel}
\usepackage{lipsum}
\usepackage{braket}
\newcommand{\zbar}{\raisebox{0.0001ex}{--}\kern-0.6em z}
\usepackage[usenames, dvipsnames]{color}
\usepackage{rotating}
\usepackage[breaklinks=true]{hyperref}
\usepackage{relsize}
\hypersetup{
	colorlinks   = true, %Colours links instead of ugly boxes
	urlcolor     = blue, %Colour for external hyperlinks
	linkcolor    = blue, %Colour of internal links
	citecolor   =  red %Colour of citations
}

\graphicspath{ {./fig/} }

\renewcommand{\H}[0]{H}  							% Hamiltonian as operator
\newcommand{\abs}[1]{\ensuremath{ \left| #1 \right| }}		% absolute value
\usepackage{epstopdf} 
\DeclareGraphicsExtensions{.pdf,.png} %png for lowres, pdf for highres

\begin{document}

\title{Bound states in microwave QED: Crossover from waveguide to cavity regime
%Waveguide QED: emergence of the infinite limit in finite systems
}

\author{N. Pradeep Kumar}
\affiliation{ARC Centre of Excellence for Engineered Quantum Systems, School of Mathematics and Physics, The University of Queensland, Saint Lucia, Queensland 4072, Australia}

\author{Andr\'es Rosario Hamann}
\affiliation{ARC Centre of Excellence for Engineered Quantum Systems, School of Mathematics and Physics, The University of Queensland, Saint Lucia, Queensland 4072, Australia}
\affiliation{Department of Physics, ETH Zürich, CH-8093 Zürich, Switzerland}

\author{Rohit Navarathna}
\affiliation{ARC Centre of Excellence for Engineered Quantum Systems, School of Mathematics and Physics, The University of Queensland, Saint Lucia, Queensland 4072, Australia}

\author{Maximilian Zanner}
\affiliation{Center for Quantum Physics, and Institute for Experimental Physics,University of Innsbruck, A-6020 Innsbruck, Austria}
\affiliation{Institute for Quantum Optics and Quantum Information of the Austrian Academy of Sciences, A-6020 Innsbruck, Austria}

\author{Mikhail Pletyukhov}
% \email{pletmikh@physik.rwth-aachen.de}
\affiliation{Institute for Theory of Statistical Physics, RWTH Aachen University, 52056 Aachen, Germany}

\author{Arkady Fedorov}
\email{a.fedorov@uq.edu.au}
\affiliation{ARC Centre of Excellence for Engineered Quantum Systems, School of Mathematics and Physics, The University of Queensland, Saint Lucia, Queensland 4072, Australia}

\date{\today}

\begin{abstract}
Light-matter interaction at the single-quantum level is the heart of many regimes of high fundamental importance to modern quantum technologies. Strong interaction of a qubit with a single photon of an electromagnetic field mode is described by the cavity/circuit electrodynamics (QED) regime which is one of the most advanced platforms for quantum computing. The opposite regime of the waveguide QED, where qubits interact with a continuum of modes in an infinite one-dimensional space, is also at the focus of recent research revealing novel quantum phenomena. Despite the demonstration of several key features of waveguide QED, the transition from an experimentally realizable finite-size system to the theoretically assumed infinite device size is neither rigorously justified nor fully understood. In this paper, we formulate a unifying theory which under a minimal set of standard approximations accounts for physical boundaries of a system in all parameter domains. Considering two qubits in a rectangular waveguide which naturally exhibits a low frequency cutoff we are able to account for infinite number of modes and obtain an accurate description of the waveguide transmission, a life-time of a qubit-photon bound state and the exchange interaction between two qubit-photon bounds states. For verification, we compare our theory to experimental data obtained for two superconducting qubits in a rectangular waveguide demonstrating how the infinite size limit of waveguide QED emerges in a finite-size system. Our theory can be straightforwardly extended to other waveguides such as the photonic crystal and coupled cavity arrays.
%Quantum emitters coupled to quantum electromagnetic field in materials with photonic bandgap provide a unique platform to study novel quantum optics phenomena such as the formation of spatially localized atom-photon bound states. We experimentally observed the formation of such bound states by coupling transmon qubits to a rectangular waveguide which has a naturally occurring low frequency cut-off.  Furthermore, by appropriately modelling the waveguide by accounting for finite length of the device as well as the metallic boundaries, we show that the interaction strength between the bound states scales exponentially with both detuning and distance of separation between the qubits. The rectangular waveguide QED system can thus serve as an equivalent 3D architecture to study bound state interactions between multiple quantum emitters.   
\end{abstract}
%\keywords{}
%\pacs{}

\maketitle

\section{Introduction}
Engineering interaction between an atom or other quantum emitters with quantized electromagnetic fields serve as the building block of modern quantum technologies. In particular, cavity quantum electrodynamics (cQED) that describes the strong coupling between an atom and a single mode of a high Q cavity is now a well understood theory and underlies the development of quantum computers based on superconducting qubits \cite{Wallraff2004, arute2019}. The natural extension of cQED leads to multimode-cQED, where several discreet modes of the cavity are simultaneously hybridized with an atom \cite{sundaresan2015,Chakram2021}. The multimode cavity satisfies the condition that $\kappa \ll \delta\omega$, where $\kappa$ is the linewidth of the modes and $\delta\omega$ is the free spectral range that measures the mode spacing \cite{PuertasMartnez2019}. In the opposite case of $\kappa \gg \delta\omega$ results in a continuum of overlapping modes that can be utilized to study waveguide-QED phenomenon that emulates the dynamics of an atom embedded in a free space \cite{Astafiev2010,abdumalikov2011}. In contrast to the standing waves of a cavity, waveguides support a large bandwidth of propagating waves that enables long-range photon mediated interaction between distant quantum emitters \cite{van2013photon,Lalumire2013}. These properties find direct applications in quantum communications for realizing quantum networks \cite{Kimble2008,Sipahigil2016}, non-reciprocal photon transmission \cite{Pichler2015,RosarioHamann2018} and routing \cite{Hoi2011}.   

Waveguides can also be engineered to support non-trivial dispersion relation that can result in photonic bandgaps or stopbands. When the qubit frequency lies in the vicinity of the band edge, a pair of dressed states are predicted to emerge. One of these dressed state merges into the continuum while the other dressed state may be pushed into the bandgap wherein the photonic part of the dressed state becomes an evanescent wave exponentially localized around the qubit \cite{bykov1975,John1990,Kofman1994,calaj2016}. Such a dressed state is referred to as the qubit-photon bound state (QPBS) and has been observed in the experiments with photonic crystals \cite{Liu2016,sundaresan2019} and metametarials \cite{scigliuzzo2021,zhang2022}. It has been proposed that the interaction between an array of such bound states can be engineered for simulating tunable spin-exchange interaction and quantum impurity models \cite{John1996,Douglas2015,shi2016, Shi2018,zhang2022}.
 
A common approach to generate stopbands is to use  photonic crystals and sub-wavelength \cite{Liu2016,sundaresan2019}  meta-materials \cite{Mirhosseini2018,scigliuzzo2021} built using an array of coupled cavities, which are typically fabricated as 2D co-planar structure on a chip that provides transverse confinement to the travelling photons. As an alternative, this can also be achieved in 3D waveguides \cite{pozar2011microwave, Shahmoon2013} which has naturally occurring low-frequency cutoff due to its geometry. Furthermore coupling between the qubits and the waveguide can be enhanced by tailoring the electric dipole of the superconducting qubit in the 3D Transmon architecture \cite{Paik2011}.

Experimental results are usually explained by postulating infinite size of a device and neglecting coupling to the input and output ports. However, physical systems employed in experiments are finite dimensional and interacts with measurement apparatus. For the passband the infinite nature is experimentally justified by using arguments of impedance matching at the coupling ports. This argument is especially  questionable for a 3D waveguide whose configuration is conceptually indistinguishable from a 3D cavity and signature of the standing waves always remain. Similarly, photonic crystals and coupled-cavity array implementations use a finite number of elementary cells and the finite nature of the system is even more pronounced. Thus, a question of validity of the infinite limit as opposed to the description which utilise the single-mode (or multi-mode) Jaynes-Cummings model \cite{PuertasMartnez2019} remains largely an open question. 

In this work we developed a unifying theory of qubits interacting with multiple number of electromagnetic modes coupled to the ports. We treat both the ports and the qubits in a similar footing, by considering the dipole type coupling with the waveguide. Furthermore, we apply the Markov approximation (only for the ports), the rotating wave approximation and considering only single-photon excitation subspace, we apply the Green's function formalism \cite{AsenjoGarcia2017,Schneider2016} which allows one to obtain analytical results for all quantities of interest. We then apply this formalism to a specific case of a rectangular 3D waveguide with two qubits where we are able to sum over infinite number of waveguide modes. We evaluate transmission through the waveguide (both empty and with the qubits), lifetime of  qubit-photon bound states in the stopband and exchange interaction of two qubit-photon bound states. Our theory shows how both the infinite limit and the opposite limit of the single-mode Jaynes-Cummings interaction emerges for a spatially finite system in different parameter domains. A particular outcome of our theory is the generalised formula for the Purcell limit and the exchange interaction between qubit which can include any (up to infinity) number of modes.

We compare our theory to experimental results for two superconducting qubits in a copper rectangular waveguide and demonstrate how the finite-size effects become negligible and the infinite waveguide limit results in an exponential scaling for bound state interaction strength. In the final part of the paper we show how our formalism can be easily extended to any arbitrary waveguide with a known dispersion relation. In particular, we discuss the case of photonic crystal where summation over infinite number of mode is not even necessary.

\section{General theory of qubit in a finite size waveguide}
Let us model a generic waveguide taking into account its finite size, the length of the waveguide along the $z$ direction is denoted by $L$. The Hamiltonian of the free waveguide can be written as, 
\begin{equation}
H_{0} = \sum_k \varepsilon_k a_k^\dag a_k \quad k = \frac{l\pi}{L}, \quad l=1,2, \dots , \label{H_0}
% \varepsilon_k &= \sqrt{\omega_c^2 + c^2 k^2}
% \label{eigenspectrum} 
\end{equation}
where $k$ is the wavevector along the $z$ direction of the waveguide and $\varepsilon_k$ is the dispersion relation associated with the waveguide, and $a_k (a_k^{\dagger})$ is the annihilation (creation) operator of the corresponding mode. Note that both the boundary conditions and the dispersion relation can be kept arbitrary at this point. 

%The eigen function of the hamiltonian are given by the following normal modes,
%\begin{equation}
%    \psi_k(z) =  \sqrt{\frac{2}{L}} \sin \, [k\, (z+L/2)],
%    \label{eigenmodes}
%\end{equation}
%satisfying the boundary conditions $\psi_k(-L/2) = \psi_k (+L/2)=0$. 

\begin{figure}[t]
    \centering
    \includegraphics[width = \columnwidth]{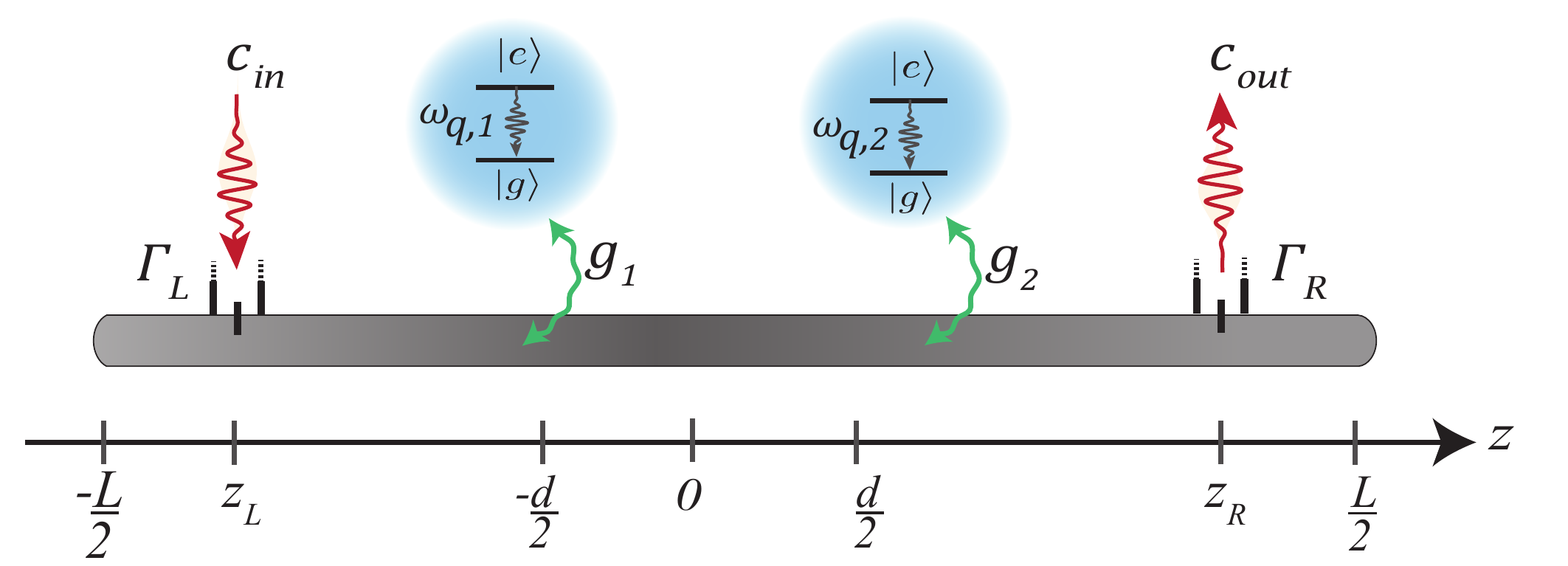}
    \caption{Schematic of a typical waveguide QED set-up employed in experiments. The waveguide is finite in size with boundaries located at $\pm L/2$. The waveguide is driven from left ($c_{in}$) to right ($c_{out}$) via input-output ports located at $z_{L,R}$ with th corresponding coupling strengths denoted by $\Gamma_{L,R}$. Pair of two-level quantum emitters ($\omega_{q,1,2}$) are coupled to the waveguide and placed symmetrically at $\pm d/2$ with respect to the origin and decays into the waveguide at the rate $g_{1,2}$.}
    \label{general_schematic}
\end{figure}

As shown in Fig.~(\ref{general_schematic}), the waveguide is coupled to the (left and right) ports at the points $z_L=-\frac{L}{2}+d_e$ and $z_R=+\frac{L}{2}-d_e$ as well as to the two qubits at the points $z_1$ and $z_2$. We describe this system by the full Hamiltonian
\begin{align}
    H = H_0 + \sum_{s=L,R} (H_s + H_{s,int}) 
    + \sum_{j=1,2} (H_j+H_{j,int}). 
    \label{H_full}
\end{align}
Here we keep the number of qubits to be two for brevity as this allows us to calculate all the single-qubit effects and the most fundamental cooperative effects such as the exchange interaction between two qubits. This configuration is also reflected in our experiment with a 3D waveguide as presented below.

Hereby the ports are modelled as structureless Markovian reservoirs with linear dispersion,
\begin{align}
    H_s = \int d \omega_s \, \omega_s \, c^{\dagger}_{\omega_s} \, c_{\omega_s} ,
\end{align}
in terms of the corresponding continuum field operators obeying the canonical commutation relations $[c_{\omega_s}, c_{\omega'_s}^{\dagger}] = \delta (\omega_s - \omega'_s)$. The ports' coupling to the waveguide is of the dipole type treated in the rotating wave approximation (RWA)
\begin{align}
    H_{s,int} =  \sum_{k} f_k a_k^{\dagger}  \sqrt{L} \psi_k (z_s) \sqrt{\frac{\Gamma_s}{\pi}} \int d \omega_s   c_{\omega_s} + h.c. ,
    \label{Hp_int}
\end{align}
where $\psi_k(z)$ is the $k$-th eigenfunction of the waveguide Hamiltonian~(\ref{H_0}). Here, we  assume that all modes of the same port are equally coupled to a given waveguide's mode $k$ --- this assumption supports the Markovian modelling of the ports. In turn, the waveguide's dipole moment may have a dispersion in the waveguide's mode index $k$. Typically this dependence is $\propto \sqrt{\varepsilon_k}$, so we define $f_k = \sqrt{\frac{\varepsilon_k}{\omega_c}}$, where $\omega_c$ is some relevant frequency scale (e.g. the low-energy cutoff frequency, like it appears in our following application).
 
The two qubits (labelled by $j=1,2$) with the transition frequencies $\omega_{q,j}$ are described by the effective Hamiltonians
\begin{align}
    H_j = \left( \omega_{q,j} - i \frac{\gamma_{a,j}}{2} \right) \frac{1+\sigma_z^{(j)}}{2},
\end{align}
which also take into account their nonradiative decay at the rates $\gamma_{a,j}$. The qubits' coupling to the waveguide is analogous to \eqref{Hp_int} --- the dipole coupling in the RWA:
\begin{align}
    H_{j,int} =  \sum_{k} f_k a_k^{\dagger}  \sqrt{L} \psi_k (z_j) g_j \sigma_-^{(j)} + h.c. ,
    \label{Hq_int}
\end{align}
where $\sigma_z^{(j)}$ and $\sigma_-^{(j)} = \frac12 [\sigma_x^{(j)} - i \sigma_y^{(j)}]$ are the standard Pauli matrices.
 
We note that the presented model is rather general and can be also applied to a lattice realization of the waveguide. In this case the sum over $k$ in \eqref{H_0} is finite, and the coordinate $z_n = n a$ is discrete and given by multiples of the lattice constant $a$.
 
\subsection{Calculating transmission}

With the help of the Hamiltonian \eqref{H_full} we derive (see details in Appendix \ref{app:eom}) the Heisenberg equations of motion for the (ports' and waveguide's) field operators and the qubits' operators. Assuming that the incident field from the left port enters the waveguide at $t=0$, we define the input field
\begin{align}
    c_{in} (t) = \int d \omega_L c_{\omega_L} (t+0^+)
    \label{c_in_t}
\end{align}
in terms of the left port operators. Since the transmitted field is 
measured in the right port, we define the output field
\begin{align}
    c_{out} (t) = \int d \omega_R c_{\omega_R} (t-0^+)
    \label{c_out_t}
\end{align}
in terms of the right port operators. The infinitesimal shifts of the time arguments in the above expressions reflect the causality property --- an important feature of the input-output formalism \cite{gardiner2004quantum}. 
The transmission amplitude $S_{RL} (\omega)$ relates the Fourier images of \eqref{c_in_t} and \eqref{c_out_t},
\begin{align}
    \tilde{c}_{out} (\omega) = S_{RL} (\omega) \tilde{c}_{in} (\omega).
    \label{input_output}
\end{align}
In the single-excitation approximation, which is justified for a weak pulse intensity and which is based on the assumption that both qubits remain close to their ground states in the time dynamics, that is $\langle \sigma_z^{(j)} (t)\rangle \approx -1$, we establish  the following relation (see again in Appendix \ref{app:eom} for more details)
\begin{align}
     S_{RL} (\omega) = -2i \sqrt{\Gamma_R \Gamma_L} L G (z_R, z_L; \omega).
     \label{SRL_G1}
\end{align}
Here 
\begin{align}
  G (z,z'; \omega) = \sum_{k,k'} f_k \psi_k (z) G_{kk'} (\omega) f_{k'} \psi_{k'} (z')  
\end{align}
is the waveguide Green's function written in the coordinate representation. Its momentum-space counterpart
\begin{align}
    G_{kk'} (\omega) = \left( \frac{1}{\omega - \hat{\varepsilon} - \Sigma (\omega) } \right)_{kk'}, \quad \hat{\varepsilon}_{kk'} = \varepsilon_k \delta_{kk'},
    \label{G_full_k}
\end{align}
is dressed by the self-energy
\begin{align}
    \Sigma_{kk'} (\omega) = \Sigma_{kk'}^{(p)} 
    + \Sigma_{kk'}^{(q)} (\omega) ,
    \label{Sigma_full_k}
\end{align}
which takes into account the effects of the interaction with the ports
\begin{align}
    \Sigma_{kk'}^{(p)} = - i L \sum_{s=L,R} \Gamma_s  f_k \psi_k (z_s) f_{k'} \psi_{k'} (z_s), 
    \label{Sigma_p}
\end{align}
and with the qubits
\begin{align}
    \Sigma_{kk'}^{(q)} (\omega) =  L \sum_{j=1,2}   g_{j}^2 \frac{  f_k \psi_k (z_j)   f_{k'} \psi_{k'} (z_j)}{\omega -  \omega_{q,j} +i \frac{\gamma_{a,j}}{2}} .
    \label{Sigma_q}
\end{align}
Note that \eqref{Sigma_p} is frequency independent, which reflects the Markovian modelling of the ports.

An alternative representation for $S_{RL} (\omega )$ can be worked out which will be applied later in Section \ref{qpbs} for the analysis of qubit-photon bound states. In particular, in order to identify these states with resonant peaks in the transmission data below the waveguide's cutoff frequency, we equivalently rewrite $S_{RL} (\omega )$ as
\begin{align}
    S_{RL} (\omega ) =& - 2 i  L  \sqrt{\Gamma_R \Gamma_L}    G^{(p)} (z_R, z_L;\omega) \label{SRL_G2a} \\
&   - 2 i  L^2  \sqrt{\Gamma_R \Gamma_L}  \sum_{j,j'}  g_j g_{j'}   G^{(p)} (z_R, z_{j} ; \omega)  \nonumber \\
& \qquad \qquad \times \hat{G}_{jj'} (\omega) G^{(p)} (z_{j'}, z_L;\omega).
    \label{SRL_G2b}
\end{align}
Here $G^{(p)} (z, z';\omega)$ is the waveguide Green's function dressed solely by the port-induced self-energy $\Sigma_{kk'}^{(p)}$: One obtains $G^{(p)} (z, z';\omega)$ by replacing $G_{kk'} \to G_{kk'}^{(p)}$ and $\Sigma_{kk'} \to \Sigma_{kk'}^{(p)}$ in \eqref{G_full_k} and \eqref{Sigma_full_k}, respectively. In addition, we introduce the dressed Green's function of the qubits,
\begin{align}
    \hat{G}_{jj'} (\omega)= \left( \frac{1}{[\hat{G}^{(0)} (\omega)]^{-1} -  \hat{\Sigma} (\omega)}\right)_{jj'},
    \label{G_qubits}
\end{align}
which is written in terms of the qubits' bare Green's function
\begin{align}
    \hat{G}_{jj'}^{(0)} (\omega) = \frac{\delta_{jj'}}{\omega - \omega_{q,j} + i \frac{\gamma_{a,j}}{2}},
    \label{G0_qubits}
\end{align}
and the qubits' self-energy
\begin{align}
    \hat{\Sigma}_{jj'} (\omega) = g_j  g_{j'}   L  G^{(p)} (z_j, z_{j'}; \omega) ,
    \label{Sigma_qubits}
\end{align}
emerging due to their interaction with the waveguide. 

Thus, the qubit-photon bound states are expected to manifest themselves as poles of \eqref{G_qubits}. We refer to Appendix \ref{app:trans_trans} for the proof of the equivalence between \eqref{SRL_G2a}, \eqref{SRL_G2b} and \eqref{SRL_G1}.

\subsection{Calculating the port-dressed Green's function of the waveguide}

Recapitulating the definition of $G_{kk'}^{(p)} (\omega)$ in terms of the Dyson equation
\begin{align}
    G_{kk'}^{(p)} (\omega) = G_{kk'}^{(0)} (\omega) + \sum_{k'', k'''} G^{(0)}_{k k''} (\omega) \Sigma_{k'' k'''}^{(p)}  G_{k''' k'}^{(p)} (\omega), 
    \label{Dyson_Gp_k}
\end{align}
where $G_{kk'}^{(0)} (\omega) = \frac{\delta_{kk'}}{\omega - \varepsilon_k}$ is the bare Green's function of the waveguide, we explicitly solve \eqref{Dyson_Gp_k} in Appendix \ref{app:dyson_SigmaP}. The obtained solution allows us to represent
(omitting for brevity the frequency argument)
\begin{align}
&  G^{(p)} (z,z') = G^{(0)} (z,z') \nonumber \\
&  -\frac{i L}{D} \Gamma_R  G^{(0)} (z,z_R)   [1+ i \Gamma_L L G^{(0)} (z_L , z_L) ] G^{(0)} (z_R,z') \nonumber \\
& -\frac{i L}{D} \Gamma_L  G^{(0)} (z,z_L)  [1+ i \Gamma_R L G^{(0)} (z_R , z_R) ]  G^{(0)} (z_L,z') \nonumber \\
&  -\frac{L^2}{D} \Gamma_R \Gamma_L G^{(0)} (z,z_R)  G^{(0)} (z_R, z_L)  G^{(0)} (z_L,z') \nonumber \\
&  -\frac{ L^2}{D} \Gamma_L  \Gamma_R  G^{(0)} (z,z_L) G^{(0)} (z_L,z_R) G^{(0)} (z_R,z'),
\label{Gp_z}
\end{align}
where 
\begin{align}
D &= [1+i\Gamma_R L G^{(0)} (z_R, z_R)][1+i\Gamma_L LG^{(0)} (z_L, z_L)]
\nonumber \\
&+L^2\Gamma_R\Gamma_L G^{(0)} (z_R, z_L)G^{(0)} (z_L, z_R),
\label{D_def}
\end{align}
and we also defined the coordinate representation of the bare waveguide Green's function
\begin{align}
    G^{(0)} (z,z') = \sum_k f_k^2 \frac{\psi_k (z) \psi_{k} (z')}{\omega - \varepsilon_k} .
    \label{G0_z}
\end{align}
It is remarkable that for the special choice of spatial arguments $z=z_R$, $z'=z_L$ the expression \eqref{Gp_z} gets considerably simplified:
\begin{align}
     G^{(p)} (z_R,z_L) = \frac{G^{(0)} (z_R,z_L)}{D}.
\end{align}

For a weak coupling to the ports we can approximate \eqref{Gp_z} by its leading order expansion in $\Gamma$'s
\begin{align}
    &  G^{(p)} (z,z') \approx G^{(0)} (z,z') \nonumber \\
    &   -i L \Gamma_R  G^{(0)} (z,z_R)   G^{(0)} (z_R,z') \nonumber \\
& -i L \Gamma_L  G^{(0)} (z,z_L)  G^{(0)} (z_L,z') ,
\label{Gp_appr}
\end{align}
which already includes the imaginary part necessary for a description of the qubit-photon bound state resonance broadening. This becomes explicit after inserting \eqref{Gp_appr} into \eqref{Sigma_qubits}.

\section{Qubits in rectangular waveguide}
\begin{figure}[t]
    \centering
    \includegraphics[width=\linewidth]{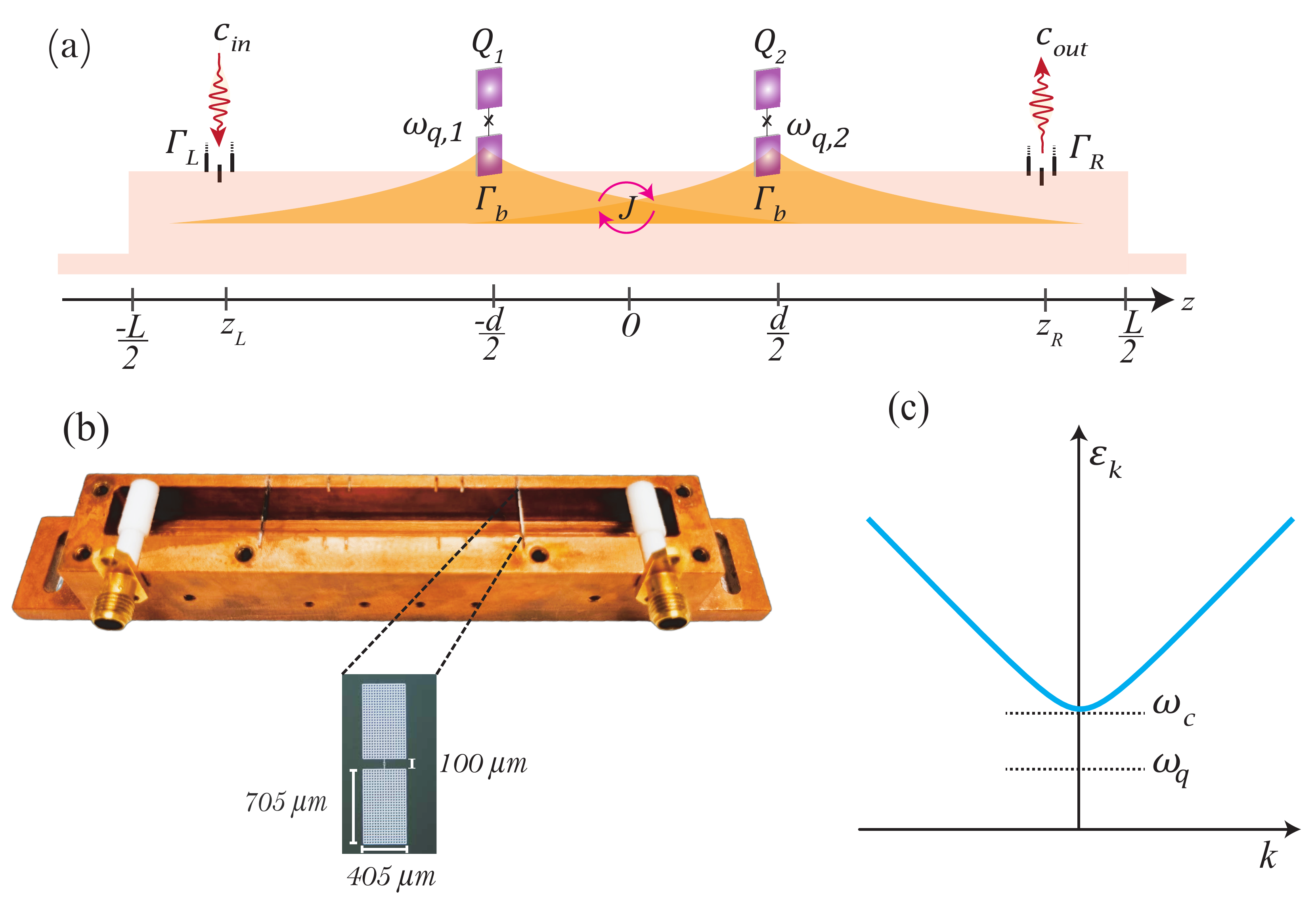}
    \caption{(a) Schematic of a rectangular waveguide containing two 3D Transmon qubits $Q_1$ and $Q_2$ separated by a distance $d$. When the qubits' frequencies lie in the stopband, exponentially localized photonic bound state can be induced, which are centered at the qubits' positions. The bound states can be probed via input-output ports separated by a distance $z_R-z_L$. (b) The actual device used in the experiment consisting of copper rectangular waveguide with co-axial connectors used for input/output ports. (c) Photon dispersion curve for the dominant $\text{TE}_{10}$ mode. The cut-off frequency $\omega_c/2\pi = 6.5213$ GHz and the qubit frequency $\omega_q \, (=\omega_{q,1} = \omega_{q,2})$ that lies in the stopband.}
    \label{schematic}
\end{figure}

Up to this point we have nowhere used the waveguide's specific properties, in particular its eigenspectrum $\varepsilon_k$ and its eigenmodes $\psi_k(z)$. Instead we have expressed all quantities which are necessary for the transmission calculation in terms of the bare Green's function of the empty waveguide \eqref{G0_z} written in the coordinate representation. In this section we demonstrate the application of our approach to model a rectangular waveguide shown in Fig.~(\ref{schematic}). It will be a subject of our subsequent  expe\-ri\-mental study in the regard of the qubit-photon bound states that arise in the stopband.
\subsection{Calculating bare Green's function of the waveguide}
The rectangular waveguide supports two types of eigenmodes, the Transverse Electric ($\text{TE}_{m,n}$) modes and Transverse Magnetic ($\text{TM}_{m,n}$) modes, with the fundamental mode being $\text{TE}_{1,0}$. For the rest of this section we assume both the ports and the qubits are predominantly coupled to the fundamental mode although coupling to $\text{TM}_{m,n}$ modes can be in principle taken into account as well. The photon dispersion associated with the $\text{TE}_{1,0}$ mode is given by \cite{pozar2011microwave}
\begin{equation}
    \varepsilon_k = \sqrt{c^2k^2+\omega_c^2}.
    \label{rect_wg_disp}
\end{equation}
The eigenfunction of the rectangular metal waveguide are given by the following normal modes, 
\begin{equation}
    \psi_k(z) =  \sqrt{\frac{2}{L}} \sin \, [k\, (z+L/2)],
     \label{eigenmodes}
\end{equation}
satisfying the open boundary conditions $\psi_k(-L/2)= \psi_k (+L/2)=0$, which are equivalent to the vanishing tangential components of the electric field ($E_x = E_y =0$) and the vanishing normal component of the magnetic field ($B_z =0$) at the endpoints of the waveguide.

% For a lattice realization of the waveguide [PRX Houck ref], the whole formalism can be employed in the present form, the finite sum in the core expression \eqref{G0_z} being easily evaluated numerically. 

Let us evaluate the bare Green's function of the empty waveguide as defined in Eq.~\eqref{G0_z}. For the continuum model of our present interest, the sum in \eqref{G0_z} includes infinitely many terms. A natural idea to truncate it at some large longitudinal mode index $N_r$ has however a flaw: The summation convergence with respect to increasing $N_r$ is very slow for the frequencies lying in the stopband. This drawback does not allow us to accurately quantify the transmission through the empty waveguide at $\omega < \omega_c$, as will be shortly illustrated.

Instead of truncating the sum in \eqref{G0_z}, we first rewrite it as
\begin{align}
    G^{(0)} (z,z'; \omega ) = \frac{\omega}{\omega_c} \sum_k \frac{\psi_k (z) \psi_{k} (z')}{\omega - \varepsilon_k}  - \frac{1}{\omega_c} \delta (z-z'),
    \label{G0_z1}
\end{align}
where the delta function emerges due the basis completeness relation $\sum_k \psi_k (z) \psi_{k} (z') = \delta (z-z')$. 

The second contribution to \eqref{G0_z1} is only efficient for $z \approx z'$, and it should be regularized on the length scale of the contact size (which has been so far treated as point-like). However, due to its frequency independence the second term can only contribute to inessential energy level renormalization effects. Therefore, we completely neglect it.

As we are interested in the transmission spectrum at frequencies $\omega \approx \omega_c$, we can also approximate the first term contributing to \eqref{G0_z2} and thus obtain
\begin{align}
    G^{(0)} (z,z'; \omega ) = \sum_k \frac{\psi_k (z) \psi_{k} (z')}{\omega - \varepsilon_k} .
    \label{G0_z2}
\end{align}
Note that the both approximations made above can be alternatively deduced from the assumption that $f_k=1$, i.e. the waveguide's dipole moment does not depend on the mode index $k$.

To calculate the infinite sum in \eqref{G0_z2}, we exploit the method frequently used in the condensed matter physics which is known as the Matsubara frequency summation \cite{Mahan2000ManyParticle}. By the analogy with the bosonic Matsubara frequencies $\omega_l = \frac{2 \pi l}{\beta}$ we define a formal analogue $\beta = \frac{2 L}{c}$  of the inverse temperature in the original method. Then we represent
\begin{align}
    c \, G^{(0)} (z,z'; \omega ) = \frac{1}{\beta} \sum_{i \omega_l} g (i \omega_l),
\end{align}
where in the conventionally denoted summation $\sum_{i \omega_l}$ the range of all integer $l$ from $-\infty$ to $+ \infty$ is implied, and $g (w)$ is a newly introduced function
\begin{align}
    g (w) =  \frac{e^{-w  \frac{|z-z'|}{c}} -  e^{- w \frac{|z+z'+L|}{c}}}{\omega -\sqrt{-w^2 +\omega_c^2}},
\end{align}
which can be analytically continued in the complex plane of $w$. It has the two branch cuts  $(-\infty , -\omega_c)$ and $(\omega_c, + \infty) $ along the real axis of $w$, as well as the two poles $w_{\pm} = \pm \sqrt{\omega_c^2 - \omega^2}\equiv \pm \bar{w}$.

The main prescription of the (bosonic) Matsubara frequency summation consists  in replacing the sum by the complex contour integral
 \begin{align}
     \frac{1}{\beta}\sum_{i\omega_l}g(i\omega_l) = \frac{1}{2\pi i} \oint dw  \frac{g(w)}{1-e^{-\beta w}},
     \label{mats_sum}
 \end{align}
 where the integration contour embraces the above mentioned singularities of the $g (w)$ function in the complex plane running in the clockwise direction. An evaluation of this integral (see in Appendix \ref{app:mats_details}) results in the following
 expression
 \begin{align}
     c \, G^{(0)} (z,z'; \omega ) =& \frac{\omega}{\bar{w}} F(z,z'; \bar{w})
     \label{pole_contrib} \\
     +& \frac{1}{\pi}  \, \int_{\omega_c}^{\infty} d \bar{\omega} \frac{\sqrt{\bar{\omega}^2 - \omega_c^2}}{\omega^2 + \bar{\omega}^2 - \omega_c^2} F (z,z'; \bar{\omega}), \label{bc_contrib}
 \end{align}
where the auxiliary function
\begin{align}
    F(z,z'; w) &=   \frac{ e^{w \frac{|z-z'|}{c}} -e^{w \frac{|z+z'+L|}{c}} }{1-e^{\beta w}} \nonumber \\
    &- \frac{ e^{-w \frac{|z-z'|}{c}} -e^{-w \frac{|z+z'+L|}{c}} }{1-e^{-\beta w}} 
    \label{F(z)}
\end{align} 
obeys the correct boundary conditions $F(\pm \frac{L}{2},z')= F (z, \pm \frac{L}{2}) =0$.

The pole contribution  \eqref{pole_contrib} to $c \, G^{(0)} (z,z'; \omega )$ is dominant. It features the exponentially decaying spatial behaviour in the stopband $\omega < \omega_c$ ($\bar{w}$ is real), and the oscillating spatial behaviour in the passband $\omega > \omega_c$ ($\bar{w}$ is imaginary). 

The branch cut contribution \eqref{bc_contrib} might be only important for $z \approx z'$ (otherwise the integrand is suppressed by quickly decaying exponential terms). So we get an approximation for the equal-point Green's function
\begin{align}
    c \, G^{(0)} (z,z; \omega ) \approx & -  \frac{\omega}{\bar{w}} - \frac{1}{\pi}  \, \int_{\omega_c}^{\omega_{hf}} d \bar{\omega} \frac{\sqrt{\bar{\omega}^2 - \omega_c^2}}{\omega^2 + \bar{\omega}^2 - \omega_c^2} \label{G_int_bc} \\
    =& -  \frac{\omega}{\bar{w}} - \frac{1}{\pi} \ln \frac{2 \omega_{hf}}{\omega_c} \\
    & + \frac{\omega}{\bar{w}} \frac{1}{\pi} \arcsin \frac{\bar{w}}{\omega_c} + O \left( \frac{\omega_c}{\omega_{hf}}\right).
    \label{G_int_bc_eval}
\end{align}
In the above integral (see in Appendix \ref{app:int_G} for details of its evaluation) we have introduced the high-frequency cutoff $\omega_{hf} \gg \omega_c$ in order to regularize its logarithmic divergence (giving thus the leading branch cut contribution which is constant in frequency). The calculated subleading contribution is nearly constant ($\approx \frac{1}{\pi}$) near the threshold $\omega \to \omega_c$ ($\bar{w} \to 0$). As it was discussed on the way from \eqref{G0_z1} to \eqref{G0_z2}, such constant contributions are negligible. 

Overall we can omit the branch cut contribution \eqref{bc_contrib} and reliably approximate $c \, G^{(0)} (z,z'; \omega)$ by the pole contribution \eqref{pole_contrib} alone. This approximation gives consistent physical results in different limiting cases considered below.

\subsubsection{Long waveguide limit}

In the long waveguide limit which is defined by the condition $\bar{w} \frac{L}{c}  \gg 1$   for $\omega<\omega_c$, the condition $| \omega_c - \omega | \gg \frac{c^2}{L^2 \omega_c}$  automatically holds and the waveguide's Green's function can be approximated by
\begin{align}
    G^{(0)} (z,z'; \omega ) \approx \frac{\omega}{c\bar{w}} F(z,z'; \bar{w}) \approx -  \frac{\omega}{c\bar{w}} e^{-\bar{w} \frac{|z-z'|}{c}},
    \label{G0_long}
\end{align}
that is by the translationally invariant expression for the infinite waveguide length, which has no signatures of the endpoints.

\subsubsection{Short waveguide limit}

For a rather short waveguide's length $L \sim \frac{c \pi}{\omega_c}$ (i.e. the longitudinal and transverse sizes of the waveguide are comparable), such that the condition $|\omega - \varepsilon_{\frac{\pi}{L}}|  \ll |\varepsilon_{ \frac{2\pi}{L}} - \varepsilon_{\frac{\pi}{L}}| \sim \frac{c \pi}{L}$ is fulfilled, we expect that the expression \eqref{G0_z2} is well approximated by the first term in the sum, i.e.
\begin{align}
G^{(0)} (z,z'; \omega)  & \approx \frac{2}{L}\frac{\sin [\frac{\pi}{L} (z+ \frac{L}{2})] \sin [\frac{\pi}{L} (z'+ \frac{L}{2})]}{\omega - \varepsilon_{\frac{\pi}{L}}} \nonumber \\
= - \frac{1}{L \bar{\delta}}  &  \left[ \cos \frac{\pi \, |z-z'|}{L} - \cos \frac{\pi \, |z+z'+L|}{L}\right],
\label{GF_1st_res}
\end{align}
where $\bar{\delta} = \varepsilon_{\frac{\pi}{L}} - \omega$. Note that $\bar{\delta}$ is defined with respect to the fundamental waveguide mode $\varepsilon_{\frac{\pi}{L}}$, not with respect to $\omega_c$.
 
Let us show that the same result can be recovered from the term \eqref{pole_contrib}. Approximating
\begin{align}
\bar{w} &= \sqrt{\omega_c^2 -\omega^2} = \sqrt{\omega_c^2 -(\varepsilon_{\frac{\pi}{L}}- \bar{\delta})^2}   \\
& \approx  \sqrt{\omega_c^2  - \varepsilon_{\frac{\pi}{L}}^2+ 2 \varepsilon_{\frac{\pi}{L}}  \bar{\delta} }
\approx  i \frac{c \pi}{L}  \left( 1 - \frac{L^2}{c^2 \pi^2}  \varepsilon_{\frac{\pi}{L}} \bar{\delta}\right) ,
\end{align}
we insert this into \eqref{pole_contrib} and obtain
\begin{align}
    & G^{(0)} (z,z'; \omega)  \approx  \frac{\omega L}{i c^2 \pi}\nonumber \\
    \times & \left[ \frac{ e^{ i \frac{\pi |z-z'|}{L}} -e^{ i \frac{ \pi |z+z'+L|}{L}} }{1-e^{-2 \pi  i  \frac{L^2}{c^2 \pi^2}  \varepsilon_{\frac{\pi}{L}} \bar{\delta} }}  - \frac{ e^{- i \frac{\pi |z-z'|}{L}} -e^{- i \frac{\pi |z+z'+L|}{L}} }{1-e^{2 \pi i   \frac{L^2}{c^2 \pi^2}  \varepsilon_{\frac{\pi}{L}} \bar{\delta} }} \right]
    \nonumber \\
    & \approx - \frac{\omega}{   L  \varepsilon_{\frac{\pi}{L}} \bar{\delta}} \left[ \cos \frac{\pi |z-z'|}{L}- \cos \frac{\pi |z+z' +L|}{L} \right].
\end{align}
To the leading order in $\bar{\delta}$, this expression is equivalent to \eqref{GF_1st_res}. The smallness condition on $\bar{\delta}$, justifying the expansions made above, reads
\begin{align}
    |\bar{\delta} | \ll \frac{c^2 \pi^2}{L^2  \varepsilon_{\frac{\pi}{L}} } 
    \sim \frac{c \pi}{L} \sim |\varepsilon_{ \frac{2\pi}{L}} - \varepsilon_{\frac{\pi}{L}}|,
\end{align}
which accords with the initially made physical assumptions for this limiting case.
 
 \subsection{Qubit-photon bound states below cutoff} \label{qpbs}
 \subsubsection{One qubit --- one bound state}
 For a single qubit $j=1$ in a 3D waveguide we obtain on the basis of Eq.~\eqref{SRL_G2b} an approximate transmission in the stopband by omitting the background empty waveguide contribution:
\begin{align}
S_{RL} (\omega) &\approx   -2 i \sqrt{\Gamma_R \Gamma_L} g_1^2 \nonumber \\
& \times \frac{L^2 G^{(p)} (z_R,z_1; \omega) G^{(p)} (z_1, z_L; \omega) }{\omega - \omega_{q,1} +i \frac{\gamma_{a,1}}{2} - g_1^2 L G^{(p)} (z_1,z_1;\omega)}.
\end{align}

Next, we approximate $G^{(p)}$ by $G^{(0)}$ in the numerator, and $G^{(p)}$ by \eqref{Gp_appr} in the denominator. The latter reads then as
\begin{align}
    & \omega - \omega_{q,1}  - \gamma \, c \, G^{(0)} (z_1,z_1;\omega) 
    \label{Re_den} \\
    +& i \left[ \frac{\gamma_{a,1}}{2} + \frac{\gamma L}{c} \sum_{s=L,R} \Gamma_s |c \, G^{(0)} (z_1 , z_s; \omega)|^2 \right],
    \label{Im_den}
\end{align}
where $\gamma = \frac{g_1^2 L}{c}$. Assuming a weak resonance broadening quantified by \eqref{Im_den}, we obtain from \eqref{Re_den} the qubit-photon bound state equation
\begin{align}
    & \omega_b - \omega_{q,1}  - \gamma \, c \, G^{(0)} (z_1,z_1;\omega_b) =0. 
    \label{bound_state} 
\end{align}

Expanding the denominator near the bound state frequency $\omega_b$, and setting $\omega = \omega_b$ in the numerator, we find the following Lorentzian approximation for the bound state resonance
\begin{align}
S_{RL} (\omega)  
& \approx  -2 i \sqrt{\Gamma_R \Gamma_L}  \frac{\gamma L}{c}  Z (\omega_b) \nonumber \\
& \times \frac{ c \, G^{(0)} (z_R,z_1; \omega_b) \, c \, G^{(0)} (z_1, z_L; \omega_b) }{\omega - \omega_{b} +i \frac{\Gamma_b}{2}} ,
\end{align}
in terms of the linewidth $\Gamma_b$ which is a sum of both non-radiative decay $\Gamma_q$ of the qubit and the radiative decay $\Gamma_{r}$, 
\begin{align}
\Gamma_b \approx 2 Z (\omega_b) \left[ \frac{\Gamma_{q,1}}{2} + \frac{\gamma L}{c} \sum_{s=L,R} \Gamma_s |c \, G^{(0)} (z_1 , z_s; \omega_b)|^2 \right]
\label{Gamma_lw}
\end{align}
and the quasiparticle weight \cite{Mahan2000ManyParticle}
\begin{align}
Z (\omega_b) = \frac{1}{1- \gamma \frac{\partial }{\partial \omega}  \, c  \, G^{(0)} (z_1,z_1;\omega)} \bigg|_{\omega \to \omega_b} < 1.
\label{Z_def}
\end{align}
For $Z \to 1$ the bound state has more weight in the qubit excited state, while for $Z \ll 1$ it is mostly photon-like (see below for a justification of this interpretation). Calculating this factor for our model, we obtain the result
\begin{align}
    Z (\omega_b) = \frac{1}{1+\frac{\gamma \omega_c^2}{(\omega_c^2 - \omega_b^2)^{3/2}} },
    \label{Z_value}
\end{align}
in which we see that the Markov approximation (consisting in the neglect of the frequency dependence in the qubits' self-energy) is justified for the bound state frequency far below $\omega_c$ (provided that $\gamma \ll \omega_c$), while for $\omega_b$ approaching $\omega_c$ the Markov approximation breaks down.

Making the Lorentzian approximation for the transmission probability 
\begin{align}
|S_{RL} (\omega)|^2 \approx \frac{A (\Gamma_b/2)^2}{(\omega - \omega_b)^2 + (\Gamma_b/2)^2},
\end{align}
we also define the amplitude at the resonance
\begin{align}
A = |S_{RL} (\omega_b)|^2 & \approx   \Big| 4  Z (\omega_b) \frac{\sqrt{\Gamma_R \Gamma_L}}{\Gamma_b} \frac{\gamma L}{c}  \Big|^2 \label{amp_theory}
 \\
& \times \Big| c \, G^{(0)} (z_1,z_R; \omega_b) \, c \, G^{(0)} (z_1, z_L; \omega_b)  \Big|^2 . \nonumber 
\end{align}

It is also worth mentioning that the representation \eqref{G_coord_sol} gives us an immediate access to the photonic part $\psi_{phot}^{(1)} (z)$ of the bound state wavefunction. Neglecting the port couplings, we identify
\begin{align}
    |\psi_{phot}^{(1)} (z)|^2 &=  \lim_{\omega \to \omega_b} [(\omega - \omega_b) G (z,z; \omega)] \\
    &= L g_1^2 Z (\omega_b) |G^{(0)} (z,z_1; \omega_b) |^2 .
    \label{phot_psi_sq}
\end{align}
Observing that
\begin{align}
   \int_{-L}^L d z |\psi_{phot}^{(1)} (z)|^2 &= - Z (\omega_b) \, \gamma \, c \frac{\partial}{\partial \omega_b} G^{(0)} (z_1,z_1 ; \omega_b) \\
   &=1-Z (\omega_b),
\end{align}
we justify the interpretation of $Z (\omega_b)$ given after its de\-finition in Eq.~\eqref{Z_def}.
In the long waveguide limit we apply the approximation \eqref{G0_long} to get
\begin{align}
    |\psi_{phot}^{(1)} (z)|^2 \propto e^{-2 \sqrt{\omega_b^2 - \omega_c^2} \frac{|z-z_1|}{c}} = e^{-2 \frac{|z-z_1|}{\xi (\omega_b)}},
\end{align}
that is $\psi_{phot} (z)$ is exponentially localized, with the localization length
\begin{align}
    \xi (\omega_b) = \frac{c}{\sqrt{\omega_c^2- \omega_b^2}},
    \label{loc_len}
\end{align} 
near the qubit position $z_1$ --- this spatial profile is actually sketched in Fig.~(\ref{schematic}a). 

\subsubsection{Two qubits --- two bound states}

In the presence of the two qubits we have two bound-state resonances which can be identified with the poles of the qubits' Green's function \eqref{G_qubits}. To neglect broadening of the resonances, we approximate $\hat{\Sigma}_{jj'} (\omega)$ in \eqref{Sigma_qubits} by
\begin{align}
    \hat{\Sigma}_{jj'} (\omega) \approx \hat{\Sigma}_{jj'}^{(0)} (\omega) = g_j g_{j'} L G^{(0)} (z_j, z_{j'}; \omega)
    \label{hat_Sigma_0}
\end{align}
as well as neglect non-radiative decay rates $\gamma_{a,j}$ in \eqref{G0_qubits}.
Then the bound states are found from the equation
\begin{align}
0 &= \det [\hat{G} (\omega)]^{-1} \approx \det \left( [\hat{G}^{(0)} (\omega)]^{-1}  - \hat{\Sigma}^{(0)} (\omega) \right) \nonumber \\
& \approx (\omega - \omega_{q,1} - \hat{\Sigma}_{11}^{(0)} (\omega)) (\omega - \omega_{q,2} - \hat{\Sigma}_{22}^{(0)} (\omega)) \nonumber \\
& -  \hat{\Sigma}_{12}^{(0)} (\omega)  \hat{\Sigma}_{21}^{(0)} (\omega).
\end{align} 
Due to the off-diagonal components $\hat{\Sigma}_{12}^{(0)} (\omega)$ and $\hat{\Sigma}_{21}^{(0)} (\omega)$ of the qubits' self-energy, which describe an effective waveguide-mediated exchange interaction between the qubits, the two bound states repel each other. The energy splitting between them quantifies the strength $J$ of the exchange interaction, and it is the subject of our next consideration.
 
For the symmetric setup with $z_1 = - z_2 = -\frac{d}{2}$ and $g_1 = g_2$, we have the identities 
$\hat{\Sigma}_{11}^{(0)} (\omega)= \hat{\Sigma}_{22}^{(0)} (\omega)$ and $\hat{\Sigma}_{12}^{(0)} (\omega)= \hat{\Sigma}_{21}^{(0)} (\omega)$.  
Tuning $\omega_{q,2}$ to the value $\omega_{q,2} = \omega_{q,1} \equiv \omega_q$ in order to have a minimally possible splitting between the two bound states, we get two separate equations for each bound state
\begin{align}
    & \omega_{b,1} - \omega_{q} - \hat{\Sigma}_{11}^{(0)} (\omega_{b,1}) -  \hat{\Sigma}_{12}^{(0)} (\omega_{b,1})=0 , \\
    & \omega_{b,2} - \omega_{q} - \hat{\Sigma}_{11}^{(0)} (\omega_{b,2}) +  \hat{\Sigma}_{12}^{(0)} (\omega_{b,2})=0 .
\end{align}

Defining the splitting $\Delta = \omega_{b,2} - \omega_{b,1}$ as well as the middle point $\bar{\omega}_b = \frac{\omega_{b,2} +\omega_{b,1}}{2}$ we derive the following self-consistent equation
\begin{align}
   \Delta = \hat{\Sigma}_{-}^{(0)} \left(\bar{\omega}_b+ \frac{\Delta}{2} \right)  - \hat{\Sigma}_{+}^{(0)} \left(\bar{\omega}_b- \frac{\Delta}{2} \right) ,
\label{self_delta} 
\end{align}
with $\hat{\Sigma}_{\pm}^{(0)} (\omega) =  \hat{\Sigma}_{11}^{(0)} (\omega) \pm  \hat{\Sigma}_{12}^{(0)} (\omega)$, which implicitly defines $\Delta$ as a function of $\bar{\omega}_b$.
 
An approximation to \eqref{self_delta} by expanding its right-hand side up to the linear order in $\Delta$ yields
\begin{align}
    \Delta (\bar{\omega}_b) & \approx  -2  \hat{\Sigma}_{12}^{(0)} (\bar{\omega}_b) Z (\bar{\omega}_b) \equiv J. \label{Delta_appr1} 
\end{align}
Alternatively, this expression can be represented (see Appendix \ref{app:split}) as
\begin{align}
    \Delta (\bar{\omega}_b) & \approx 2 | \langle \psi_{phot}^{(1)} | (\bar{\omega}_b -H_0 ) |  \psi_{phot}^{(2)} \rangle | ,
    \label{J_overlap}
\end{align}
that is (two times) the overlap between the photonic contributions $\psi_{phot}^{(1)}$ and $\psi_{phot}^{(2)}$ to the corresponding bound states at their degeneracy point $\bar{\omega}_b$, subtracting the correction term $\langle \psi_{phot}^{(1)} | H_0 |  \psi_{phot}^{(2)} \rangle$ to avoid the double-counting of $H_0$ (since it is used each time for determining both $\psi_{phot}^{(1)}$ and $\psi_{phot}^{(2)}$ independently of each other).

In the long waveguide limit $L \gg \xi (\bar{\omega}_b)$  the coupling between the two qubits is approximated by 
\begin{align}
    J & \approx   2 \gamma  \frac{\bar{\omega}_b \, \xi (\bar{\omega}_b)}{c} e^{- \frac{d}{\xi (\bar{\omega}_b)}}  \frac{1}{1+ \gamma \frac{\omega_c^2}{c^3} \xi^3 (\bar{\omega}_b)}.
    \label{Delta_appr2}
\end{align}

In the single-mode limit we obtain the coupling between the two qubits by
approximating \eqref{Delta_appr1} using \eqref{GF_1st_res} and \eqref{hat_Sigma_0}:
\begin{align}
    J \approx  2 Z (\bar{\omega}_b) \frac{\tilde{g}_1 \tilde{g}_2}{\varepsilon_{\frac{\pi}{L}} -\bar{\omega}_b} ,
\end{align}
where
\begin{align}
\tilde{g}_j = g_j \, \sqrt{L} \psi_{\frac{\pi}{L}} (z_j) ,
\label{gj_tilde}
\end{align}
and
\begin{align}
Z (\bar{\omega}_b) & = \left[ 1+\frac{\tilde{g}_j^2}{( \varepsilon_{\frac{\pi}{L}}-\bar{\omega}_b)^2} \right]^{-1}.
\label{Z_single_mode}
\end{align}
In particular, $Z (\bar{\omega}_b) \approx 1$ for $|\varepsilon_{\frac{\pi}{L}} -\bar{\omega}_b| \gg | \tilde{g}_j |$.

\begin{figure}[t]
    \centering
    \includegraphics[width = \columnwidth]{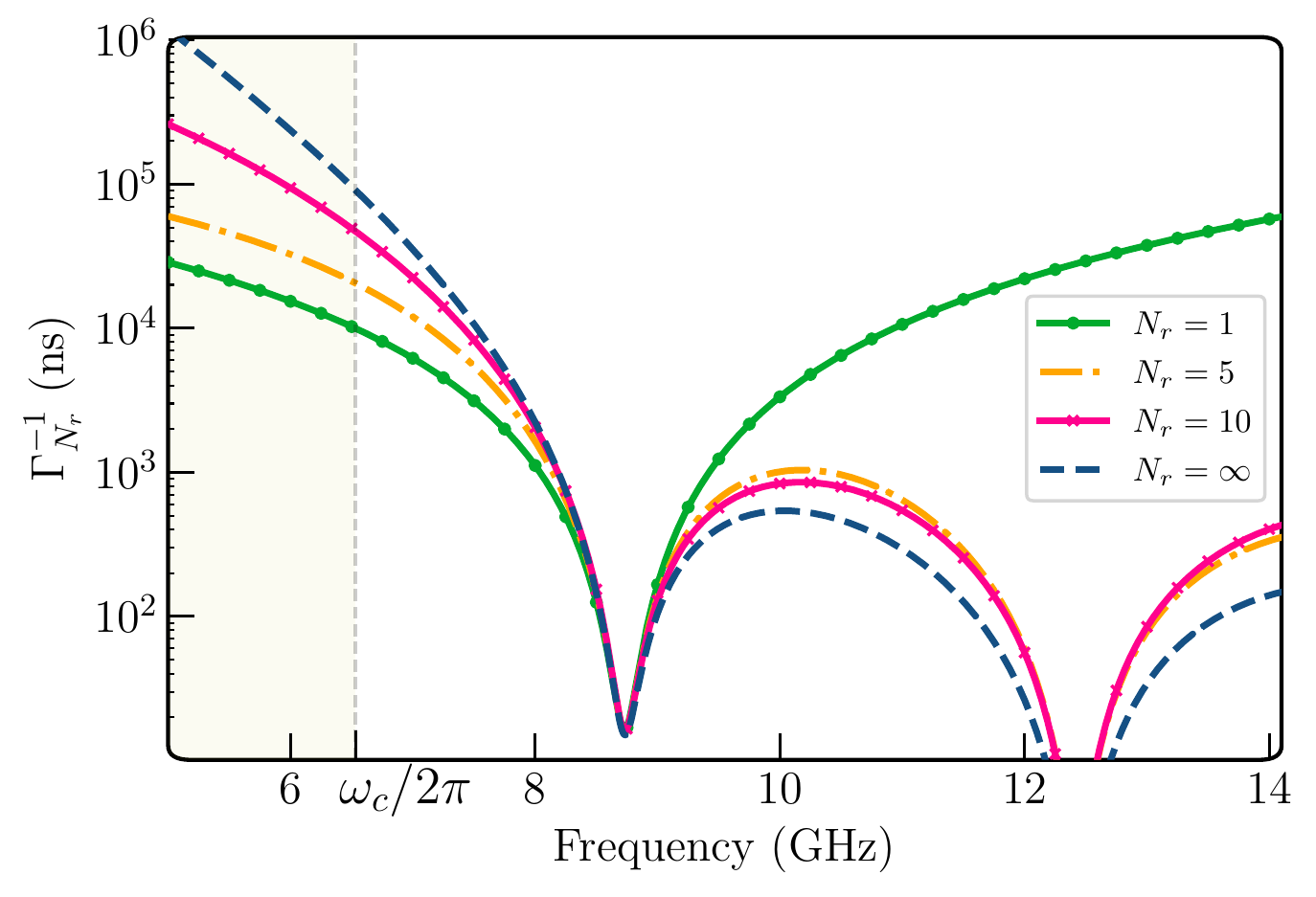}
    \caption{Comparison of the Purcell decay rate of a qubit coupled to a cavity with different number of modes $N_r$.}
    \label{fig:purcell_plot}
\end{figure}

\section{Generalized Purcell effect}
The spontaneous emission rate of a qubit depends on both the nature of the electromagnetic environment to which it is coupled as well as the detuning between the frequency of the qubit and the modes of the environment. In our formalism, the most general formula for calculating the Purcell induced decay is given by, 
\begin{equation}
    \Gamma_{r,N_r \rightarrow \infty} = 2 Z (\omega_b) \frac{\gamma L}{c}  \sum_{s=L,R} \Gamma_s |c \, G^{(0)} (z_1 , z_s; \omega_b)|^2 .
    \label{gen_purcell_formula}
\end{equation} 
The Green's function $G^{(0)}$ carries the information about the environment as seen by the qubit. For a rectangular waveguide that supports the infinite number of modes, the  Green's function is given by Eqs.~(\ref{pole_contrib}), (\ref{bc_contrib}).

Similarly, for a qubit coupled to a multimode environment with a finite number of resonances $N_r$, such as a photonic crystal and a cavity array with $N=N_r$ cavities,  the Green's function is represented as a discrete sum (see an exemplary evaluation in the framework of the tight-binding model in Appendix \ref{app:lattG}). The decay rate can be then estimated by the following formula
\begin{equation}
    \Gamma_{r,N_r} = 2 Z (\omega_b) \frac{\gamma L}{c} \sum_{s=L,R} \Gamma_s \abs{c \sum_{l=1}^{N_r} \frac{\psi_{k_l} (z_1) \psi_{k_l} (z_s)}{\omega_b - \varepsilon_{k_l}}}^2 .
\end{equation}

In the case of a single mode cavity ($N_r=1$), where the qubit in only coupled to the fundamental resonance $l=1$, we obtain the bound state decay rate by approximating Eq.~(\ref{gen_purcell_formula}) with Eq.~(\ref{GF_1st_res}),
\begin{align}
\Gamma_{r,N_r=1} & \approx Z (\omega_b) \frac{\tilde{g}_j^2}{ (\varepsilon_{\frac{\pi}{L}} -\omega_b)^2} \, \kappa, 
\label{Purcell} 
\end{align}
where $\tilde{g}_j$ is defined in \eqref{gj_tilde}, and
\begin{align}
\kappa &= \sum_{s=L,R} 2 \Gamma_s  \,  | \sqrt{L} \psi_{\frac{\pi}{L}} (z_s) |^2
\end{align}
is the cavity decay rate. Taking into account \eqref{Z_single_mode}, which holds in the single-mode case,
we observe that near $\varepsilon_{\frac{\pi}{L}} = \omega_b$ the decay rate \eqref{Purcell} is regularized,
\begin{align}
\Gamma_{r,N_r=1} & \approx \frac{\tilde{g}_j^2}{ (\varepsilon_{\frac{\pi}{L}} -\omega_b)^2 + \tilde{g}_j^2} \, \kappa, 
\label{Purcell2} 
\end{align}
and coincides with the cavity decay rate $\kappa$ at $\varepsilon_{\frac{\pi}{L}} = \omega_b$. In turn, for $|\varepsilon_{\frac{\pi}{L}} - \omega_b| \gg |\tilde{g}_j|$ it holds $Z (\omega_b) \approx 1$, and we obtain  the well known formula  \cite{houck2008} for Purcell induced decay
\begin{align}
\Gamma_{r,N_r=1} & \approx \frac{\tilde{g}_j^2}{ (\varepsilon_{\frac{\pi}{L}} -\omega_b)^2} \, \kappa .
\label{Purcell3} 
\end{align}
In Fig.~(\ref{fig:purcell_plot}) we have compared the life time of a qubit coupled to a cavity with different number of modes. Below the fundamental frequency, the propagating modes are suppressed and the qubit predominantly couples to evanescent modes. Therefore, in the limit of $N_r \rightarrow \infty$, the qubit has the largest lifetime. In contrast, for frequencies above the fundamental mode the qubit life is modified due to presence of higher harmonic modes which contributes to the density of states available at the qubit frequency. Consequently the qubit life time decreases with increase in the number of modes. 

\section{Experiment}
    \begin{figure}[t]
        \centering
        \includegraphics[width = \columnwidth]{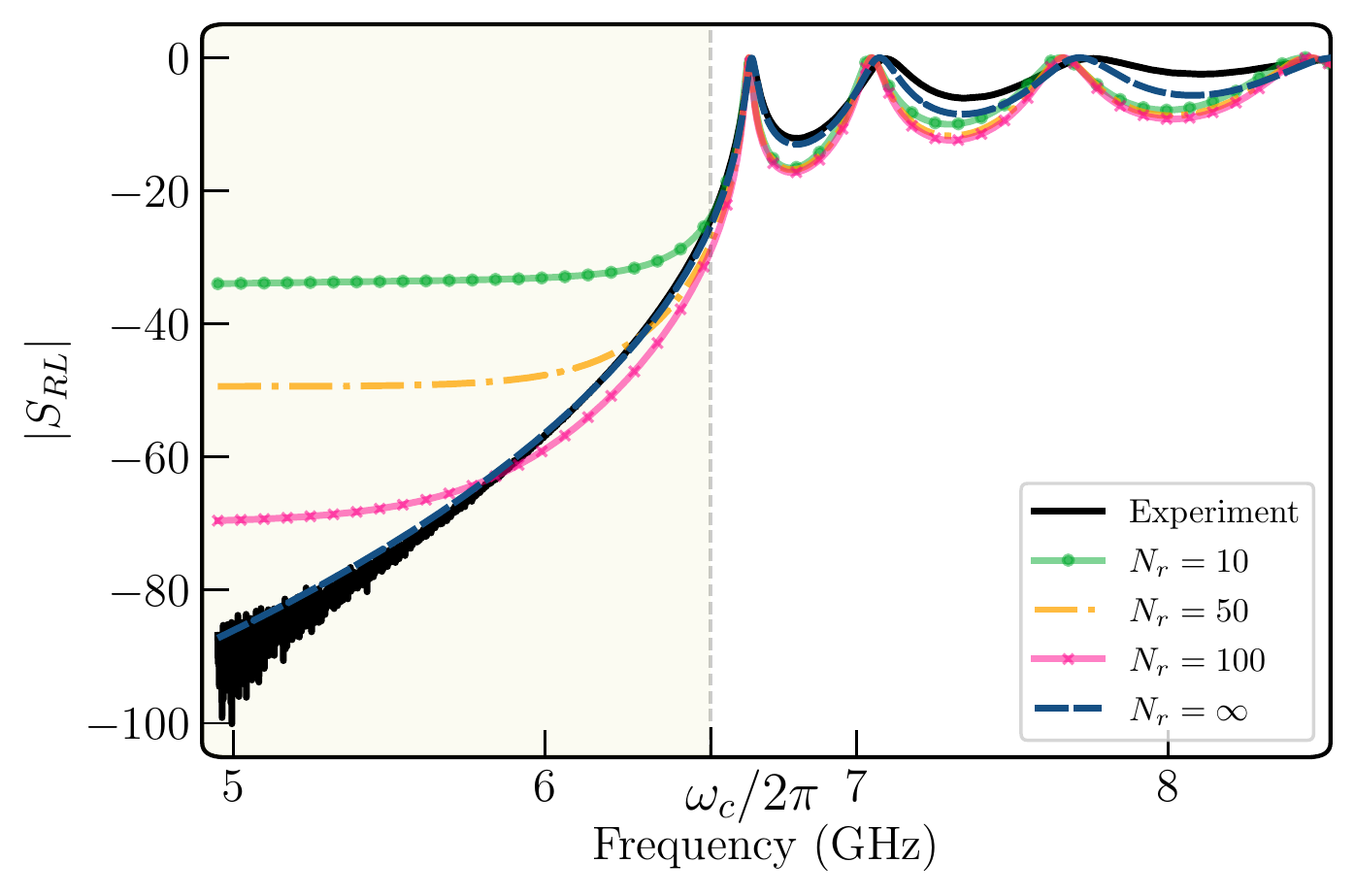}
        \caption{Transmission of the empty waveguide measured at room temperature (black solid line). It is compared to the theoretical results computed using two methods. The dot-dashed lines result from  the numerical simulations of $S_{RL}$ using the truncated sums in \eqref{G0_z2} at $N_r = 10, \ 50 \ \text{and} \ 100$ resonances. The dashed blue line is calculated using the exactly evaluated sum in \eqref{G0_z2} (i.e. at $N_r \rightarrow \infty$). Exploiting its good agreement with the experimental data, we estimate on the basis of its analytical expression the values of  $\Gamma_{L,R}/2\pi \approx 225$ MHz. The cut-off frequency is at $\omega_c/2\pi=6.5213$ GHz.}
        \label{fig:emptywg_trans}
    \end{figure}

\subsection{Waveguide characterization}
    We first characterize the waveguide without qubits through a transmission measurement at room temperature and fit the experimental data to theoretical values given by Eq.~(\ref{SRL_G2a}). Note that the coupling between the waveguide and the ports $\Gamma_{s=L, R}$ are the only free parameters in the fit. We also let $\Gamma_L = \Gamma_R = \Gamma_{L,R}$, which is a reasonable assumption since we use the identical co-axial connectors for both the ports. Furthermore, we observe that the transmission amplitude reaches unity in the passband of the waveguide indicating that there is no left-right asymmetry in the couplings. In Fig.~(\ref{fig:emptywg_trans}) we have compared the transmission data to two different theoretical results. The first one is based on the approximation where the waveguide's Green's function is represented by a sum of a finite number of resonances $N_r$ truncating the sum in Eq.~(\ref{G0_z2}) --- effectively this is equivalent to replacing the continuum model by a lattice counterpart with $N_r$ sites and the lattice constant $L/N_r$. In turn, the full sum ($N_r \rightarrow \infty$)  provides the exact description of the continuum model leading to the analytic expression Eq.~(\ref{pole_contrib}) for the (bare) Green's function. We clearly see that the continuum limit best represents the data and provides us with the estimate of $\Gamma_{L,R}/2\pi \approx 225$ MHz. We also note that the small mismatch in fitting the passband can be caused by our neglect of the weak frequency dependence of the coupling between the waveguide modes and the ports which we made in deriving the formula for the transmission. However, the minor effect of this couplings' dispersion is inessential in the stopband, where all subsequent measurements are performed. \\
    \begin{figure*}[t]
        \centering
          \includegraphics[width = \textwidth]{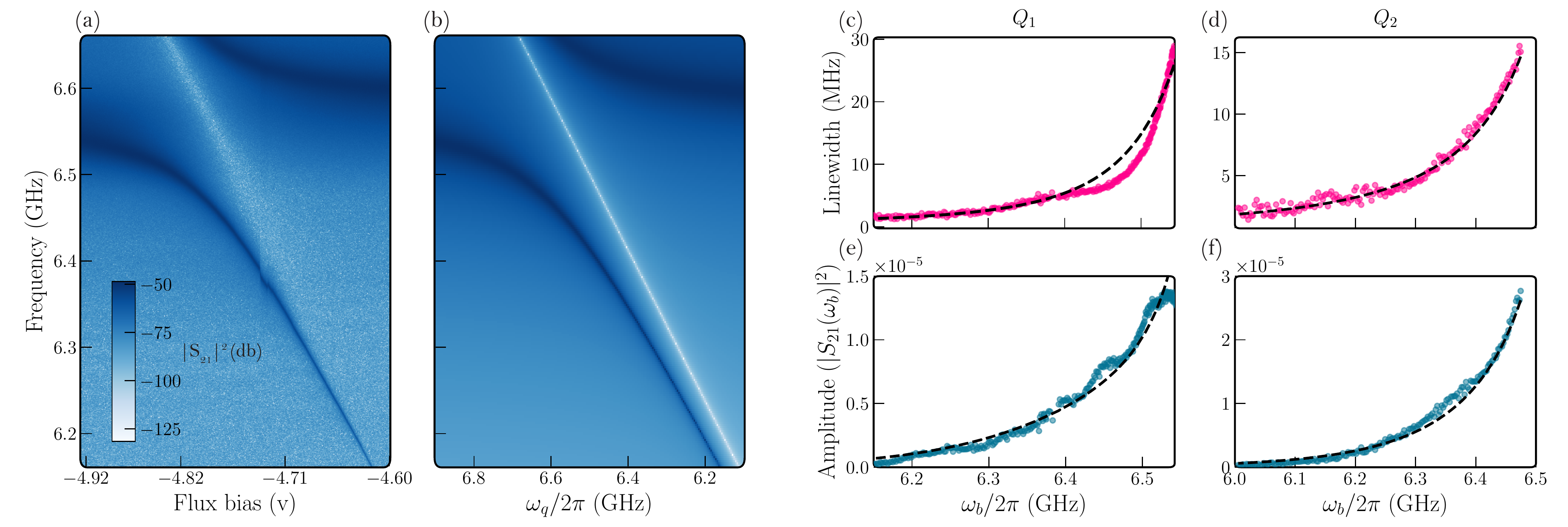}
          \caption{(a) Measurement data and (b) is the simulation of waveguide spectroscopy reveals the dressed QPBS appearing as a peak in the stopband. The frequency of the QPBS is shifted lower by a self energy $\Sigma_{11}$ from the qubit frequencies, which in turn appears as a dip. Note that the bound state always exists in the stopband, for qubit frequencies close to the cutoff it is photon-like ($Z \approx 0$) and for frequencies deep in the stopband it is qubit-like ($Z \approx 1$). (c) and (d) are linewidth of the bound state measured from spectroscopy data for qubits $Q_1$ and $Q_2$ placed symmetrically from the center of the waveguide. (e) and (f) shows the attenuation of peak amplitude as a function of bound state frequency which has similar trend as that of linewidth. The only free parameter in the fitting is $\Gamma_{L,R}/2\pi \approx 525$ MHz.}
          \label{qpbs_spec}
    \end{figure*} 
    
\subsection{Probing QPBS in the stopband}
    In order to detect QPBS in our experiment, we performed a spectroscopy measurement by sweeping the frequency $\omega_{q,1}$ of the first qubit ($Q_1$) through the cutoff into the stopband of the waveguide and observed the scattering properties. Through out the experiment the frequency $\omega_{q,2}$ of the second qubit ($Q_2$) is placed in the passband and hence far detuned from $\omega_{q,1}$. As shown in Fig.~(\ref{qpbs_spec}a), when we tune the bare qubit frequency $\omega_{q,1}$ close to $\omega_c$, we observe a peak in transmission signal in the stopband as a result of the formation of a single photon dressed bound state that is localized around  the position $z_1$ of the qubit and extends to the ends of the wave\-guide. The localization length $\xi$ given by \eqref{loc_len} depends on the bound state frequency $\omega_b$, which differs from the qubit frequency by the self-energy term, $\omega_{b} - \omega_{q,1} = \hat{\Sigma}_{11}(\omega_{b}) \approx \hat{\Sigma}_{11}(\omega_{q,1})$. Thus, $\xi$ is tuned {\it in situ} by changing $\omega_{q,1}$. 
    
    %Furthermore, the frequency of the bound state differ from the qubit frequency by a self-energy  which can be calculated just using the bare Green's function of the waveguide as shown in Eq.~(\ref{hat_Sigma_0}). 
    
    Although the density of states vanishes in the stopband, the bound state acquires a finite linewidth due to the coupling to the input and output ports. The strength of this coupling depends on $\xi$, therefore measuring linewidth of the bound state serves as an indirect measure of the localization length. We extracted linewidth as a function of the bound state frequency $\omega_{b}$ for both $Q_1$ and $Q_2$ from independent spectroscopy measurements. In Figs.~(\ref{qpbs_spec}a,b), we compared experimental data with our theoretical estimate given in Eq.~(\ref{Gamma_lw}) and find good agreement for both the qubits. Since we measure the output of the waveguide through the right port, the extent of the bound state induced by $Q_1$ which is close to the port is large compared to the bound state induced by $Q_2$. This results in smaller linewidth for the bound state localized around $Q_2$ which can be observed by comparing Fig.~(\ref{qpbs_spec}a) and (\ref{qpbs_spec}b). We also observe that when the frequency of the bound states are $\omega_{b}/2\pi \approx 6.1$ GHz, the linewidth has the lowest value of $\Gamma_b/2\pi \approx 1$ MHz for both the bound states, which is approximately close to the bare linewidth ($\Gamma_q$) of the qubits. This indicates that in the span of $\approx 400$ MHz from $\omega_c$ the bound states are completely localized in our system. 
    
    Similar to the linewidth, the transmission amplitude of the bound state resonance also depends on $\omega_b$. In Figs. (\ref{qpbs_spec}e,f)  we show the attenuation of the bound state signal as function of the frequency and find a good agreement between the theoretical prediction (see Eq.~(\ref{amp_theory})) and the data for both $Q_1$ and $Q_2$. The qubit-waveguide coupling strength $g_{1,2}$ needed for the fitting is obtained from the avoided-crossing data which will be discussed in section (\ref{qpbs_J}). Therefore the fit obtained for both linewidth and amplitude has no free parameters which shows the validity of the theory for our experimental results. 
\subsection{QPBS as an effective two-level system}
    %Rabi measurement
    Inducing a localized photonic mode in the bandgap can also be achieved by introducing dislocations in the unit cell of a photonic crystal. This principle underlies the working of defect mode lasers where the localized mode acts as a high finesse cavity~\cite{yablonovitch1987, painter1999}. In contrast, QPBS acts as a non-linear defect as it inherits the anharmonic nature of the bare qubit seeding it. In order to probe the internal structure of the bound state, we detuned the qubit by $450$ MHz below the cutoff frequency such that the bound state frequency is at $\omega_b/2\pi = 6.077$ GHz and the corresponding linewidth is measured to be $\Gamma_b/2\pi = 0.143$ MHz. When the qubit frequency lies deep in the stopband one can make the Markovian approximation to eliminate the waveguide modes and treat the bound state as an effective two-level system with finite anharmonicity (see Appendix \ref{app:dynamics}).
    
    When probed with a microwave tone, the bound state responds by emitting photon out of both the input $c_{in}$ and the output $c_{out}$ ports. Since the input signal is strongly attenuated in the stopband, measurement of the radiation coming out of the output port $\langle c_{out}\rangle$ will be a direct measure of the bound state polarization $\langle \sigma_-^{(1)}\rangle$ as shown in Eq.~(\ref{in_out_eff2}). This should be contrasted to the measurement in the passband where the radiation will be a sum of the incoming radiation and the qubit emission described in Eq.~(\ref{in_out_eff1.1}) and (\ref{in_out_eff1.2}). The large spacial separation between the ports exponentially suppresses the coupling through the waveguide modes and minimizes direct cross-talk between the ports enabling direct measurement of the bound state polarization even in the presence of the probe field.
    
\begin{figure}[t]
    \centering
        \includegraphics[width=\columnwidth]{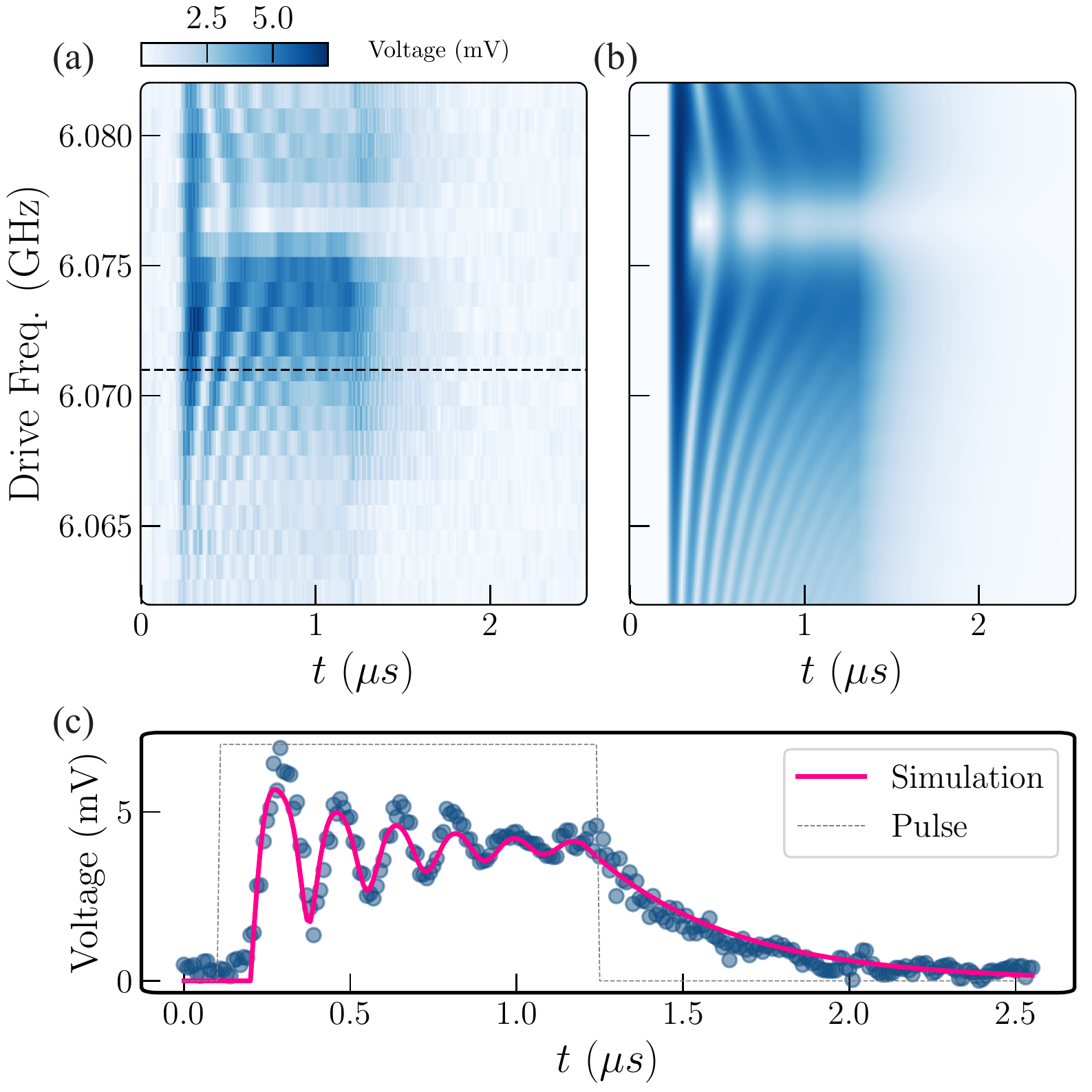}
        \caption{(a) Chevron pattern of the Rabi oscillations obtained by sweeping drive frequency in the vicinity of the bound state frequency $\omega_b/(2\pi) = 6.077$ GHz. The length of the drive pulse is $\tau_p = 1.3 \ \mu s$ during which we observe oscillations. The signal decays at the rate $\propto 2\Gamma_b$ after the pulse is turned off. (b) Numerical simulation of the Chevron pattern with $\Gamma_b/\Omega_r\sim 0.28$ reproduces qualitatively the features in the experimental data. (c) Horizontal line-cut (orange dashed line) of (a). The signal recorded here is proportional to direct emission form the bound state.}
        \label{Rabi_chevron}
    \end{figure}

    To elucidate the qubit nature of the QBPS we apply an excitation pulse of fixed length $\tau_p = 1.3 \ \mu s$ to input port while measuring the emission from the output port. In Fig.~\ref{Rabi_chevron}(a) one can see the emission as function of time and driving frequency $\omega_d$ exhibiting the characteristic chevron pattern of Rabi oscillations (at frequency $\Omega_r$) for the pulse duration. More specifically, Fig.~\ref{Rabi_chevron}(b) shows a horizontal line cut of which shows decaying Rabi oscillations for the duration of the pulse $\tau_p$ and decays exponential to zero after that. We also notice that, when the drive frequency becomes resonant with the bound state frequency $\omega_b$, the Rabi oscillations vanishes and the emission has a constant value for the time period $\tau_p$ of the drive. The amplitude of this emission depends the drive strength, dissipation rate and the detuning $\delta_d = \omega_d - \omega_b$ as shown in Eq.~(\ref{app:ss_dyanmics}). The steady state thus obtained is not unique to bound states but a simple feature of a driven-dissipate two level system which cannot be reveled in a typical circuit-QED setup due to the absence of direct coupling to $\langle \sigma_- \rangle$ component of the qubit. The numerical simulation of the Lindblad master equation for the qubit reduced density matrix $\frac{d \rho_q}{dt} =-i [H_{\text{eff}}, \rho_q] + \Gamma_b \left( \sigma_-^{(1)} \rho_q \sigma_+^{(1)}-\frac12 \{ \sigma_+^{(1)} \sigma_-^{(1)},\rho_q \}\right)$, with $H_{\text{eff}} = -\frac{\delta_d}{2} \sigma_z^{(1)} + \frac{\Omega_r}{2} \sigma_x^{(1)}$, which is equivalent to (\ref{app:master_eq1}) and (\ref{app:master_eq2}) upon identifying $\langle \hat{\sigma}^{(1)}_{\pm,z} (t) \rangle = \text{Tr} [\sigma^{(1)}_{\pm,z} \rho_q (t)]$, reproduces all features that we observe in the experimental data as shown in Fig.~(\ref{Rabi_chevron}).
    \begin{figure*}[t]
        \centering
        \includegraphics[width=\textwidth]{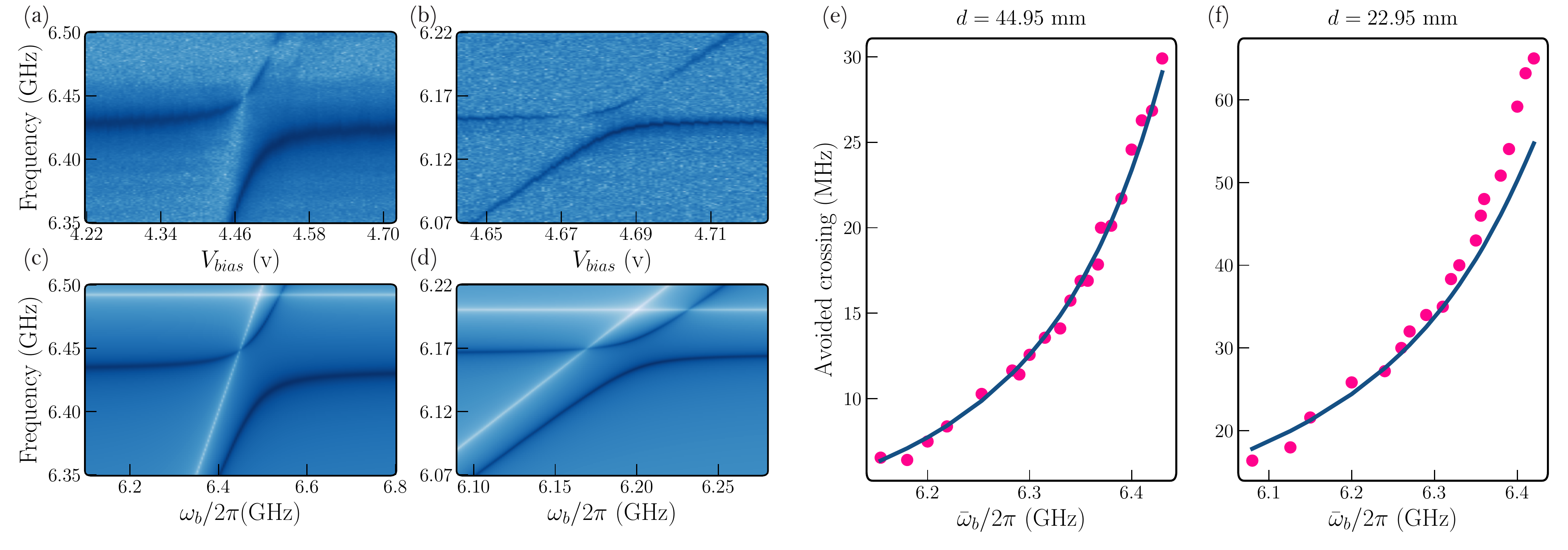}
        \caption{(a) Avoided crossing measured by fixing qubit $\omega_{q_1}$ and sweeping qubit $\omega_{q_2}$, when both the bound states are on resonance the coupling is $J=60$ MHz. (b) Same as (a) but the bound state is further detuned from the cutoff frequency, here $J = 15$ MHz. (c) and (d) are the simulation of the spectroscopy experiment that reproduces all the features, the dip in transmission happens when the bound states comes into resonance with the qubit. (e) and (f) shows the avoided crossing measured at different bare bound state frequencies $\bar{\omega}_b$ for two different distance of separation between the qubits. For $d=44.95$ mm, the formula obtained from the infinite waveguide limit gives an accurate description of the avoided crossings. We use this the   data to estimate the qubit-waveguide coupling $g_1/2\pi = g_2/2\pi$ to be $62.53$ MHz. When the inter-qubit distance is reduced to $d=22.95$ mm we notice that the approximate formula starts to breaks down for avoided crossing measured close to the cutoff.}
        \label{AC_spec}
    \end{figure*}  
\subsection{Non-radiative dipole-dipole interaction between two QPBS}\label{qpbs_J}
    %Avoided crossing
    So far in our experiment, we have considered the formation of a single QPBS centered at the qubit co-ordinate when $\omega_q < \omega_c$. However, many interesting quantum effects in waveguide QED arise due to distinctive photon mediated interaction between multiple quantum emitters. In this section, we study the interaction between bound states that can be established by allowing the individual photonic wavefunctions to overlap. Such type of interaction can be described by transverse field exchange Hamiltonian of the form $H_I = \frac12 J [\sigma_+^{(1)} \sigma_-^{(2)} + h.c.]$. In order to measure the coupling $J$ between the qubits we tuned their transition frequencies into the stopband such that two bound states are induced at the qubits' positions. The amount of overlap between the two bound states depends on the detuning between the bound state frequencies and the cut-off frequency. When both the bound states are on resonance a single virtual excitation is exchanged between them leading to the formation of symmetric ($\ket{+}$) and anti-symmetric ($\ket{-}$) pair of states: $\ket{\pm} = (\ket{ge} \pm \ket{eg})/\sqrt{2}$, where $\ket{g}$ and $\ket{e}$ are the ground and first exited levels of the bound states. We measured the coupling $J$ with respect to the "bare" bound state frequency $\tilde{\omega}_b=(\omega_{b,1}+\omega_{b,2})/2$ through spectroscopy measurements shown in Fig.~(\ref{AC_spec}). When detuning $\delta/2\pi = (\omega_c - \tilde{\omega}_b)/2 \pi$ is $50$ MHz, the measured coupling strength $J/2\pi = 66$ MHz as shown in Fig.~(\ref{AC_spec}a). In comparison, the size of the avoided crossing reduces to $J/2\pi = 15$ MHz when the qubit is tuned deep into the stopband where the detuning $\delta/2\pi = 15$ MHz as shown in Fig.~(\ref{AC_spec}b). We also utilized the general two qubit transmission model described in Eq.~(\ref{SRL_G2b}) and reproduced all the features that we observe in the experiment as shown in Fig.~(\ref{AC_spec}a \& b). 
    
    The coupling between two QPBS can be shown (see Eq.~(\ref{Delta_appr2})) to decay exponentially with $J \propto e^{-|z_{1} - z_{2}|/\xi}$ in the infinite-waveguide limit (applicable for $L \gg \xi$), where $\xi$ is the {\it in situ} tunable localization length, see Eq.~(\ref{loc_len}). This is in contrast to interactions mediated by propagating modes in passband which is oscillatory. In order to show the tunable nature of this interaction, we have measured the size of the avoided crossings when both the qubits are on-resonance as a function $\tilde{\omega}_b$. Note that the interaction strength not only depend on detuning but also on the distance $d$ of separation between the qubits. As shown in Fig. (\ref{AC_spec}), we measured the coupling for two different distances and observed that $J$ scales exponentially as predicted by Eq.~(\ref{Delta_appr2}). Furthermore, by comparing Fig.~(\ref{AC_spec}(e \& f) we precisely find when the infinite waveguide limit is a good approximation. In our system, when the inter-qubit distance is large $d=44.95$ mm we find that infinite limit gives the accurate description of the avoided crossings. On the other hand, it breaks down when the inter-qubit distance is small $d=22.95$ mm. The only free parameters in the fitting are the qubit-waveguide couplings $g_{1,2}$, we assume that both the qubits have approximately same coupling strength $g_1 \approx g_2 = g$ and estimated it to be $g/2\pi \approx 62.53$ MHz which we have also used in fitting Fig.~(\ref{qpbs_spec}). It is important to note that although the exponential nature of the interaction may indicate that this interaction would vanish in the long waveguide limit, $\xi$ can be made equal to the length of the total system by appropriately choosing $\delta$ and $d$, hence the interaction is effectively long-range. 
    % \begin{figure}[t]
    %     \centering
    %     \includegraphics[width=\columnwidth]{Figures/Avoided_crossing_sweep.pdf}
    %     \caption{Avoided crossing measured at different bare bound state frequencies $\bar{\omega}_b$ for two different distance of separation between the qubits. (a) For $d=44.95$ mm, the formula obtained from the infinite waveguide limit gives an accurate description of the avoided crossings. We use this the   data to estimate the qubit-waveguide coupling $g_1/2\pi = g_2/2\pi$ to be $62.53$ MHz. (b) When the inter-qubit distance is reduced to $d=22.95$ mm we notice that the approximate formula starts to breaks down for avoided crossing measured close the cutoff.}
    %     \label{AC_sweep}
    % \end{figure}  
    
\section{Summary and conclusions}
    In summary we developed a theory which allows us to calculate all the parameters of interest for a system of two qubits coupled to a waveguide of finite length.
    More specifically, knowing the waveguide dispersion relation $\varepsilon_k$ and its wavefunctions $\psi_k(z)$ for a given waveguide length $L$ it is possible to account for boundaries through the calculation of the waveguide Green's function given by Eq.~($\ref{G0_z}$). Our formulae then cover all parameter regimes as well as properties of the qubits and waveguide above and below cutoff.
    
    For the systems like 1D arrays of coupled cavities or 1D photonic crystals the Green's function can be found by direct summation analytically (for a few cavities/cells) or numerically (for a large number of cavities/cells). For the 3D waveguide we take into account an infinite number of modes analytically by Eqs.~(\ref{pole_contrib})-(\ref{F(z)}).
    
    Using the qubit-photon states below cutoff as a benchmark we show that the infinite limit arises when $\sqrt{\omega_c^2-\omega^2}L/c\gg1$. Given that the localization length of the photonic part of a wavefunction of a bound state is $\xi=c/\sqrt{\omega_c^2-\omega^2}$ we have a simple physical interpretation: the infinite limit holds as far as the photonic wave-functions of the qubit-photon bound state does not reach the boundaries of the waveguide. In this limit the exponential spacial decay of the waveguide Green's function arises~(\ref{G0_long}) which is then casted into exponential localization of the photonic part of the qubit-photon bound state and the experimentally observed exponential dependence of various parameters on the qubits detuning.
    
    It is also interesting to note that in the infinite limit the ports are completely decouple from the problem and neither validity of the infinite limit nor the physics below cutoff depends on the coupling strength to the ports. Thus, the infinite limit will apply even for the case of the small coupling where the transmission in the passband will not be flat and will show sharp resonance transmission peaks.
    
    Eqs.~(\ref{pole_contrib}-\ref{F(z)}) can also reproduce the Jaynes-Cummings model behaviour for the opposite short waveguide limit.  However, it should be noted that that once the detuning of the qubit from the fundamental mode increases the short cavity limit will break down and more modes will be taken into account. It is especially visible for our generalised Purcell formula (\ref{gen_purcell_formula}) describing the life-time of the qubit-bound state deviate from the conventional Purcell formula (\ref{Purcell}) as the qubit detuning grows.
    We compare our theoretical results to the experimental data where we coupled two superconducting transmon qubits to a rectangular copper waveguide. Here we focused on the regime where the qubit frequencies are tuned below the cutoff frequency in the stopband of the 3D waveguide. In this regime, both the linewidth and the amplitude of the transmission peaks corresponding to the qubit-photon bound states below cutoff as well as the exchange interaction strength between the two qubit-photon bound states show exponential dependence as predicted by the infinite limit formulae. 
    
    Overall, our results provide an important step for waveguide QED regime with the missing rigorous justification of the infinite-size limit for finite-size physical implementations. Furthermore, out 3D waveguide QED system serve as an alternative platform for studying interacting qubit-photon bound states. In the future work a high fidelity dispersive readout of the bound state can be performed by inserting a planar readout resonator inside the waveguide\cite{zoepfl2017characterization}, thereby effectively realizing circuit-QED in our set-up. We have also shown the tunable nature of interaction between two bound states through long range non-radiative dipole-dipole interaction. The exponential nature of these interaction can be potentially used for performing quantum simulations of quantum spin-models in a 3D-architecture \cite{dalmonte2015realizing}.

\begin{acknowledgements}
We thank E. Shahmoon for fruitful discussions. The authors were supported by the Australian Research Council Centre of Excellence for Engineered Quantum Systems (EQUS, CE170100009). 
\end{acknowledgements}

\begin{appendix}

\section{Heisenberg equations of motion and derivation of the expression \eqref{SRL_G1} for the transmission}
\label{app:eom}

With the help of the Hamiltonian \eqref{H_full} we derive the Heisenberg equations of motion
\begin{align}
\frac{d}{d t} a_k (t)&= -i \varepsilon_k a_k (t) \nonumber \\
&-i \sqrt{L} \sum_{s=L,R} f_k \psi_k (z_s) \sqrt{\frac{\Gamma_s}{\pi}} \int d \omega_s c_{\omega_s} (t)    \nonumber  \\
&- i\sqrt{L} \sum_{j=1,2} f_k \psi_k (z_j)  g_{j} \sigma_-^{(j)}  (t) , \label{eq_dak} \\
\frac{d}{d t} c_{\omega_s} (t) &= -i \omega_s c_{\omega_s} (t) \nonumber \\
&- i  \sqrt{\frac{\Gamma_s}{\pi}} \sum_{k} \sqrt{L} f_k \psi_k (z_s)  a_k (t), \label{eq_c} \\
\frac{d}{dt} \sigma_-^{(j)} (t) &= -i \left( \omega_{q,j} -i \frac{\gamma_{a,j}}{2} \right) \sigma_-^{(j)} (t) \nonumber \\
&+ i \sum_k  g_{j} \sqrt{L} f_k \psi_ k (z_j) \sigma_z^{(j)} (t) a_k (t). \label{eq_Q} 
\end{align}

Integrating the port field operators
\begin{align}
c_{\omega_s} (t)  &= c_{\omega_s} (0)e^{-i \omega_s t} \\
&- i \sqrt{\frac{\Gamma_s}{\pi}} \sum_{k} \sqrt{L} f_k \psi_k (z_s) \int_{0}^{t} dt' e^{-i \omega_s (t-t')} a_k (t'), \nonumber
\end{align}
we next evaluate
\begin{align}
 \int d \omega_R c_{\omega_R} (t)  & =  -  i \sqrt{\frac{\Gamma_R}{\pi}}  \int d \omega_R \sum_{k} \sqrt{L} f_k \psi_k (z_R) \nonumber \\
 & \times \int_{0}^{t} dt' e^{-i \omega_R (t-t')} a_k (t') \\
&=   - 2  i \sqrt{\pi \Gamma_R}  \sum_{k} \sqrt{L} f_k \psi_k (z_R) \nonumber \\ 
& \times \int_{0}^{t} dt' \delta (t-t') a_k (t')  \\
&=- i \sqrt{\pi \Gamma_R } \sum_{k} \sqrt{L} f_k \psi_k (z_R) a_k (t) .
\label{cR_time} 
\end{align}
Note that we used $\int_0^t dt' \delta (t-t') = \frac12$, and set $c_{\omega_R} (0)=0$, which means no input pulse in the right port. In turn,
\begin{align}
\int d \omega_L c_{\omega_L} (t)  & = \int d \omega_L c_{\omega_L} (0) e^{-i \omega_L t} \nonumber \\
&- i \sqrt{\pi \Gamma_L } \sum_{k} \sqrt{L} f_k \psi_k (z_L) a_k (t) .
\label{cL_time}
\end{align}

For the input field we analogously obtain
\begin{align}
c_{in} (t) &=\int d \omega_L  c_{\omega_L} (t + 0^+)  \\
&=  \int d \omega_L  c_{\omega_L} (0)e^{-i \omega_L (t+0^+)}  \nonumber \\
&- i  2 \pi  \sqrt{\frac{\Gamma_L}{\pi}} \sum_{k} \sqrt{L} f_k \psi_k (z_L)
\nonumber \\
& \times \int_{0}^{t} dt' \delta (t + 0^+ -t') a_k (t') .
\end{align}
The last term vanishes, since $t' = t + 0^+$ lies outside the integration range, and thus obtain
\begin{align}
c_{in} (t) =  \int d \omega_L  c_{\omega_L} (0)e^{-i \omega_L t}.
\end{align}

Next, we define the output field
\begin{align}
c_{out} (t) &= \int d \omega_R  c_{\omega_R} (t - 0^+)  \\
&=  - i  2 \pi  \sqrt{\frac{\Gamma_R}{\pi}} \sum_{k} \sqrt{L} f_k \psi_k (z_R) \nonumber \\
& \times \int_{0}^{t} dt' \delta (t - 0^+ -t') a_k (t') \\
&=  - i  2  \sqrt{\pi \Gamma_R} \sum_{k} \sqrt{L} f_k \psi_k (z_R)   a_k (t).
\label{out_a_exact}
\end{align}
Note that in contrast to \eqref{cR_time} we get now the twice larger contribution, since $t' = t - 0^+ < t$ entirely lies inside the integration range.

Performing the Fourier transform
\begin{align}
\tilde{a}_k (\omega) = \int_0^{\infty} d t a_k (t) e^{i (\omega +i 0^+) t}, 
\end{align}
we obtain
\begin{align}
\tilde{c}_{out} (\omega)  = -2 i    \sqrt{\pi \Gamma_R} \sum_{k} \sqrt{L} f_k \psi_k (z_R)   \tilde{a}_k (\omega).
\label{out_field}
\end{align}

Inserting the solutions \eqref{cR_time} and \eqref{cL_time} into \eqref{eq_dak}, we obtain
\begin{align}
\frac{d}{d t} a_k (t) &= -i \varepsilon_k a_k (t) - i \sum_{k'} \Sigma_{kk'}^{(p)} a_{k'} (t)    \\
&- i  \sum_{j=1,2} \sqrt{L} f_k \psi_k (z_j)  g_{j} \sigma_-^{(j)} (t)  \\
&- i \sqrt{L} f_k \psi_k (z_L) \sqrt{\frac{\Gamma_L}{\pi}}  c_{in} (t),
\end{align}
where the port-induced self-energy $\Sigma_{kk'}^{(p)}$ is defined in \eqref{Sigma_p}. In the Fourier representation this equation reads
\begin{align}
 & (\omega -\varepsilon_k) \tilde{a}_k (\omega) - \sum_{k'} \Sigma_{kk'}^{(p)} \tilde{a}_{k'} (\omega) \nonumber \\
 &-  \sqrt{L} \sum_{j=1,2} f_k \psi_k (z_j) g_{j} \tilde{\sigma}_-^{(j)} (\omega)  \nonumber \\
& =  \sqrt{\frac{\Gamma_L}{\pi}} \sqrt{L} f_k \psi_k (z_L) \tilde{c}_{in} (\omega) .
\label{a_om1}
\end{align}

In the equation  \eqref{eq_Q} we approximate $\sigma_z^{(j)} (t) \approx -1$, which is justified for a weak power of the incident pulse. Thus we get the equations
\begin{align}
\frac{d}{dt} \sigma_-^{(j)} (t) &= -i \left( \omega_{q,j} -i \frac{\gamma_{a,j}}{2} \right) \sigma_-^{(j)} (t) \\
&- i  \sum_k  g_{j} \sqrt{L} f_k \psi_k (z_j) a_k (t). \label{eq_Qa}  
\end{align}
Rewriting it in the Fourier representation,
we eliminate $ \tilde{\sigma}_-^{(j)} (\omega)$:
\begin{align}
 \tilde{\sigma}_-^{(j)}  (\omega) &=  \frac{1}{\omega -  \omega_{q,j} +i \frac{\gamma_{a,j}}{2}}  \sum_k  g_{j} \sqrt{L} f_k \psi_k (z_j) \tilde{a}_k (\omega).
\end{align}
Inserting these relations into \eqref{a_om1} yields
\begin{align}
 &  \sum_{k'}[ (\omega -\varepsilon_k) \delta_{kk'}  - \Sigma_{kk'} (\omega) ] \tilde{a}_{k'} (\omega) \nonumber \\
 &=  \sqrt{\frac{\Gamma_L}{\pi}} \sqrt{L} f_k \psi_k (z_L) \tilde{c}_{in} (\omega) ,
 \label{a_eq}
\end{align}
where  $\Sigma_{kk'} (\omega)$ is the full self-energy of the waveguide defined in \eqref{Sigma_full_k}.

Introducing the fully dressed waveguide Green's function \eqref{G_full_k},
we solve \eqref{a_eq} for $\tilde{a}_{k} (\omega)$:
\begin{align}
\tilde{a}_{k} (\omega)  =  \sqrt{\frac{\Gamma_L}{\pi} }  \sum_{k'} G_{kk'} (\omega) \sqrt{L} f_{k'}  \psi_{k'} (z_L) \tilde{c}_{in} (\omega) .
\end{align}
Inserting the result into \eqref{out_field}, we obtain the relation \eqref{input_output} between the input and the output fields in terms of the transmission amplitude 
\begin{align}
S_{RL} (\omega) &= - 2 i  L  \sqrt{\Gamma_R \Gamma_L} \sum_{k,k'}  f_{k} \psi_k (z_R)   G_{kk'} (\omega)  f_{k'} \psi_{k'} (z_L) , \nonumber 
\end{align}
which accords with \eqref{SRL_G1}.

\section{Transmission representation \eqref{SRL_G2a}, \eqref{SRL_G2b}}
\label{app:trans_trans}

To achieve the representation \eqref{SRL_G2a}, \eqref{SRL_G2b} for the transmission amplitude $S_{RL} (\omega)$, we attempt to solve the Dyson equation for the fully dressed Green's function $G_{kk'}$ of the waveguide
\begin{align}
G_{kk'} &= G_{kk'}^{(p)} + \sum_{k'',k'''} G^{(p)}_{k k''} \Sigma_{k'' k'''}^{(q)}  G_{k''' k'} \label{G_Dyson} \\
&\equiv G_{kk'}^{(p)} + \sum_{j,j'} \bar{Q}^{(p)}_{k j} \hat{G}_{jj'}^{(0)} Q_{j' k'}   .
\label{GQ}
\end{align} 
Hereby the port-dressed Green's function $G_{kk'}^{(p)}$ of the waveguide is defined 
\begin{align}
G^{(p)}_{kk'} (\omega) = \left( \frac{1}{\omega - \hat{\varepsilon} - \Sigma^{(p)}} \right)_{kk'}
\label{Gp}
\end{align}
in terms of the port-induced self-energy \eqref{Sigma_p}. The qubit-induced self-energy $\Sigma_{kk'}^{(q)}$ is given in \eqref{Sigma_q}. In addition, we define $ \tilde{\psi}_{k} (z)= f_k  \psi_{k} (z)$,
\begin{align}
\bar{Q}^{(p)}_{kj}  &= \sum_{k''}  G^{(p)}_{k k''} \sqrt{L}   \tilde{\psi}_{k''} (z_j) g_j , \\
Q_{j' k'}^{(p)}  &= g_{j'} \sum_{k''}  \sqrt{L} \tilde{\psi}_{k''} (z_{j'}) G_{k'' k'}^{(p)},
\end{align}
and the analogous quantities without the superscript $(p)$. Finally, the bare qubits' Green's function $\hat{G}^{(0)}_{jj'}$ is given in \eqref{G0_qubits}.

Appropriately convoluting \eqref{G_Dyson}, we arrive at the equation
\begin{align}
Q_{j'k'}  = Q_{j'k'}^{(p)} + \sum_{j,j''} \hat{\Sigma}_{j' j}   \hat{G}_{jj''}^{(0)} Q_{j'' k'}   ,
\label{Q_eq}
\end{align}
where $\hat{\Sigma}_{jj'} $ is the full qubits' self-energy given in \eqref{Sigma_qubits}.

Inverting the matrix in the qubits' space in \eqref{Q_eq}, we obtain
\begin{align}
 Q_{jk'}   = \sum_{j'} \left( \frac{1}{1-\hat{\Sigma} \hat{G}^{(0)}} \right)_{jj'} Q_{j'k'}^{(p)} .
\end{align} 
Inserting this expression into \eqref{GQ}, we establish
\begin{align}
G_{kk'}  &= G_{kk'}^{(p)} + \sum_{j,j'} \bar{Q}^{(p)}_{k j}   \left( \hat{G}^{(0)} \frac{1}{1-\hat{\Sigma} \hat{G}^{(0)}} \right)_{jj'} Q_{j'k'}^{(p)} \\
&= G_{kk'}^{(p)} + \sum_{j,j'} \bar{Q}^{(p)}_{k j}   \hat{G}_{jj'} Q_{j'k'}^{(p)} ,
\label{Gsol}
\end{align}
where $\hat{G}_{jj'}$ is the fully dressed qubits' Green's function given in \eqref{G_qubits}. Hence, we find the coordinate-space solution
\begin{align}
    G (z,z'; \omega) &=  G^{(p)} (z,z'; \omega) \nonumber \\
    &+  L \sum_{j,j'}  g_j g_{j'}   G^{(p)} (z, z_{j}; \omega) \nonumber \\
    & \qquad \times \hat{G}_{jj'} (\omega) G^{(p)} (z_{j'} , z'; \omega) ,
    \label{G_coord_sol}
\end{align}
which immediately leads to the representation \eqref{SRL_G2a}, \eqref{SRL_G2b} of the transmission amplitude \eqref{SRL_G1}.

\section{Solution of the Dyson equation \eqref{Dyson_Gp_k}}
\label{app:dyson_SigmaP}

We notice that the Eq.~\eqref{Dyson_Gp_k} is analogous to Eq. \eqref{G_Dyson}: The latter is obtain from the former by the replacements
\begin{align}
& G_{kk'} \to G_{kk'}^{(p)}, \\
& G_{kk'}^{(p)} \to G^{(0)}_{k k'}, \\
& \hat{G}_{jj'}^{(0)} \to - i \pi \delta_{jj'}, \quad j,j' \to L,R , \\
& g_j \to \sqrt{\frac{\Gamma_j}{\pi}}, \\
& \hat{\Sigma}_{jj'} \to \frac{\sqrt{\Gamma_j \Gamma_{j'}}}{\pi} L G^{(0)} (z_j, z_{j'}).
\end{align}
Therefore we can exploit the solution \eqref{G_coord_sol}, making in it the same replacements. Most of them are obvious, and we comment only on the matrix $\hat{G}_{jj'}$: In the $(L,R)$ basis it reads
\begin{align}
    &\hat{G} =\nonumber \\
    &\left( \begin{array}{cc} -\frac{1}{i \pi} -\frac{\Gamma_L}{\pi} L G^{(0)} (z_L, z_L) & -\frac{\sqrt{\Gamma_L \Gamma_R}}{\pi} L G^{(0)} (z_L, z_R) \\ -\frac{\sqrt{\Gamma_R \Gamma_L}}{\pi} L G^{(0)} (z_R, z_L) & -\frac{1}{i \pi} -\frac{\Gamma_R}{\pi} L G^{(0)} (z_R, z_R)\end{array} \right)^{-1} \nonumber \\
    &= - \frac{i \pi}{D} \\
    & \!\! \times \left( \begin{array}{cc} 1+i L \Gamma_R G^{(0)} (z_R , z_R) & -i L \sqrt{\Gamma_L \Gamma_R} G^{(0)} (z_L , z_R) \\ -i L \sqrt{\Gamma_R \Gamma_L} G^{(0)} (z_R , z_L) & 1+i L \Gamma_L G^{(0)} (z_L , z_L) \end{array} \right)  \nonumber 
\end{align}
together with $D$ defined in \eqref{D_def}. Expanding the sums over $j$ and $j'$ in \eqref{G_coord_sol} we obtain \eqref{Gp_z}.

\section{Further details on evaluation of the Matsubara sum \eqref{mats_sum}}
\label{app:mats_details}

\begin{figure}[t]
  \centering
  \includegraphics[width = \columnwidth]{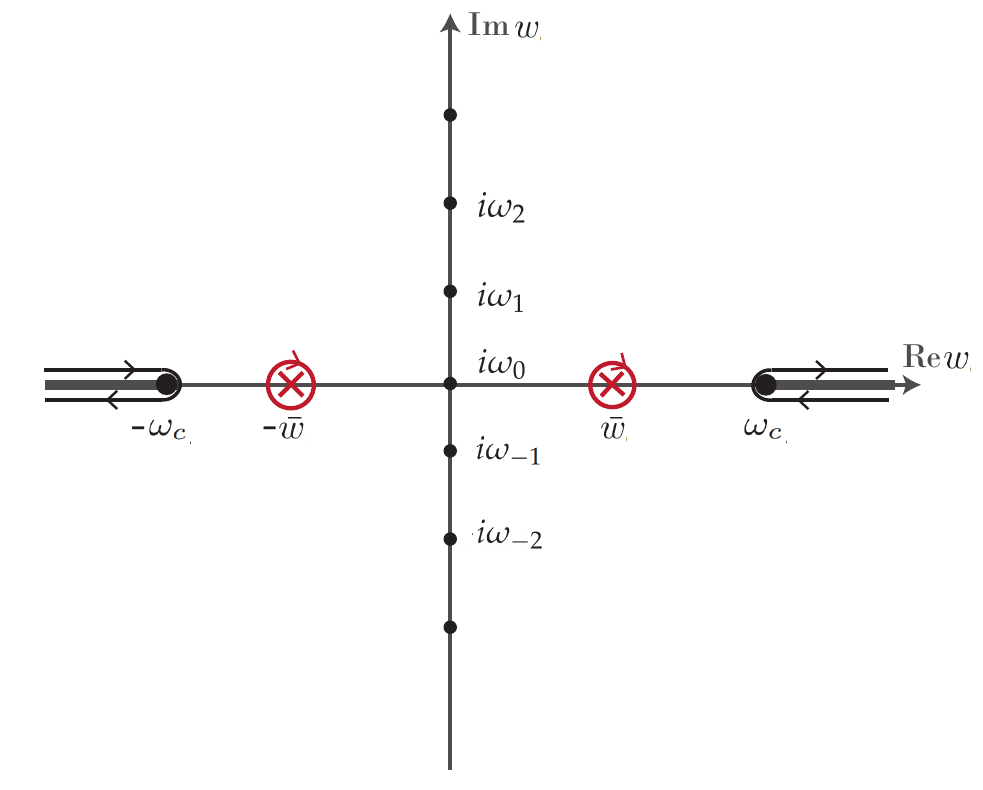}
  \caption{Integration contour for the integral in the right-hand side of \eqref{mats_sum} in the complex plane of $w$. It embraces the poles $w= \pm \bar{w}$ of $g (w)$ (the red crosses) as well as the branch cuts of $g (w)$ (the branching points $w=\pm \omega_c$ are indicated by the fat black points).}
  \label{contour}
\end{figure}

To switch from the sum in the left-hand side of \eqref{mats_sum} to the integral in its right-hand side, which is performed along the contours shown in Fig.~(\ref{contour}), it is also necessary to make sure that at $\text{Re} \, w \neq 0$ the integrand goes to zero,
\begin{align}
    \Big| \frac{g (w)}{1-e^{-\beta w}} \Big| \to 0 ,
\end{align}
faster than $\frac{1}{|w|}$ as $|w| \to \infty$. 

Excluding the special point $z=z'$ (which we consider separately later), we first inspect the case $\text{Re}\, w >0$ and observe the exponentially decaying terms $e^{-|\text{Re}\,  w| \frac{|z-z'|}{c}}$, $e^{-|\text{Re} \, w| \frac{|z+z'+L|}{c}}$.

For $\text{Re} \, w <0$ we obtain the exponentially decaying terms $e^{-|\text{Re}\,  w| (\frac{2L}{c}-\frac{|z-z'|}{c})}$, $e^{-|\text{Re} \, w| (\frac{2L}{c} -\frac{|z+z'+L|}{c})}$, provided that $z$ and $z'$ do not appear on the waveguide's endpoints. The latter condition is however not restrictive, since we know that the bare Green's function vanishes if one of its arguments approaches $+ \frac{L}{2}$ or $-\frac{L}{2}$.

Having established the sufficient conditions to deform the integration contour to the shape shown in Fig.~(\ref{contour}), we easily find the pole contribution \eqref{pole_contrib} by evaluating the corresponding residua values. In turn, to evaluate the branch cut integrals we observe that
\begin{align}
    & g (\bar{\omega}+ i 0^+) - g (\bar{\omega}-i 0^+) = (e^{-\bar{\omega}  \frac{|x-x'|}{c}} -  e^{- \bar{\omega} \frac{|x+x'+L|}{c}})\nonumber \\ & \times \left[\frac{1}{\omega +i  \, \text{sgn} (\bar{\omega}) \sqrt{\bar{\omega}^2 -\omega_c^2}}  -  \frac{1}{\omega -i  \, \text{sgn} (\bar{\omega}) \sqrt{\bar{\omega}^2 - \omega_c^2}}\right] .
\end{align}
This observation leads to the expression \eqref{bc_contrib}.

In the special case $z=z'$, the sum in the left-hand side of \eqref{mats_sum} diverges logarithmically, and we regularize it by subtracting from it the analogous sum with $\omega =0$. Defining the function
\begin{align}
    \tilde{g} (w) &= \frac{1}{\omega - \sqrt{-w^2 +\omega_c^2}} + \frac{1}{ \sqrt{-w^2 +\omega_c^2}} \\
    &=  \frac{\omega}{ \sqrt{-w^2 +\omega_c^2}} \frac{1}{\omega - \sqrt{-w^2 +\omega_c^2}},
\end{align}
we see that it produces the same pole contribution \eqref{pole_contrib} as $g (w)$ does at $z=z'$ (when neglecting the regular contribution $\propto  e^{-w \frac{|z+z'+L|}{c}}$). Eventually this observation means that the proposed regularization is equivalent to introducing the high-frequency cutoff $\omega_{hf}$ in \eqref{G_int_bc}, which also captures the logarithmically diverging frequency-independent contribution.

\section{Evaluation of the integral in \eqref{G_int_bc}}
\label{app:int_G}

Let us perform the integral in \eqref{G_int_bc} by making the variable change $\bar{\omega} = \omega_c \cosh \lambda$:
\begin{align}
      &- \frac{1}{\pi}  \, \int_{\omega_c}^{\omega_{hf}} d \bar{\omega} \frac{\sqrt{\bar{\omega}^2 - \omega_c^2}}{\omega^2 + \bar{\omega}^2 - \omega_c^2} \nonumber \\
      &=    - \frac{1}{\pi}  \, \int_{0}^{\text{arccosh} \frac{\omega_{hf}}{\omega_c}} d \lambda \frac{ \sinh^2 \lambda}{(\frac{\omega}{\omega_c})^2  + \sinh^2 \lambda} \\
      & \approx - \frac{1}{\pi} \ln \frac{2 \omega_{hf}}{\omega_c} + \frac{1}{\pi} \frac{\omega^2}{\omega_c^2} \int_0^{\infty} \frac{2 d\lambda}{2 (\frac{\omega}{\omega_c})^2  -1 + \cosh 2 \lambda},
      \label{f_cosh}
\end{align}
where the last approximation is valid for $\omega_{hf} \gg \omega_c$.

The next change of variable $y= e^{2\lambda}$ leads to
\begin{align}
    & \int_0^{\infty} \frac{2 d\lambda}{2 (\frac{\omega}{\omega_c})^2  -1 + \cosh 2\lambda} \nonumber \\
    &= \int_{-\infty}^{\infty} \frac{d\lambda}{2 (\frac{\omega}{\omega_c})^2  -1 + \cosh 2\lambda} \nonumber \\
    &=\int_0^{\infty} \frac{dy}{(y+ \cos 2 \alpha)^2 + \sin^2 2 \alpha} \nonumber  \\
    &=  \frac{2 \alpha}{\sin 2 \alpha} = \frac{\omega_c^2}{\omega \bar{w}} \arcsin \frac{\bar{w}}{\omega_c} ,
    \label{f_arc}
\end{align}
where we have parameterized $\omega= \omega_c \cos \alpha$ and $\bar{w}= \sqrt{\omega_c^2 - \omega^2} = \omega_c \sin \alpha$. Combining \eqref{f_arc} with \eqref{f_cosh}, we obtain the result shown in \eqref{G_int_bc_eval}.

\section{Explanation of the formula \eqref{J_overlap}}
\label{app:split}

Traditionally, the bound state in the presence of a single qubit is found from the eigenvalue problem $H |\psi_{BS}^{(1)} \rangle = \omega_b | \psi_{BS}^{(1)} \rangle$, where in the Hamiltonian \eqref{H_full} the coupling to the ports and to the second qubit is neglected. Representing
\begin{align}
    |\psi_{BS}^{(1)} \rangle = \alpha^{(1)} |eg\rangle |\text{vac}\rangle + |gg\rangle \sum_k \beta_k^{(1)} a_k^{\dagger}|\text{vac}\rangle , 
\end{align}
where $|\text{vac}\rangle$ is the photonic vacuum state, we obtain the following equations for the coefficients $\alpha^{(1)}$ and $\beta_k^{(1)}$:
\begin{align}
    (\omega_b - \omega_{q,1}) \alpha^{(1)} &= g_1 \sum_k \sqrt{L} \psi_k (z_1)  \beta_k^{(1)} , \\
    (\omega_b - \varepsilon_{k}) \beta_k^{(1)} &= g_1 \sqrt{L} \psi_k (z_1)  \alpha^{(1)} .
\end{align}
It follows
\begin{align}
    |\psi_{BS}^{(1)} \rangle = \alpha^{(1)} \left[ |eg\rangle |\text{vac}\rangle + |gg\rangle \sum_k \frac{g_1 \sqrt{L} \psi_k (z_1)}{\omega_b - \varepsilon_{k}} a_k^{\dagger}|\text{vac}\rangle \right].
    \label{BS1}
\end{align}
In particular, from this expression we recover the photonic part \eqref{phot_psi_sq} of the bound state wavefunction
\begin{align}
    \psi_{phot}^{(1)} (z) &= \alpha^{(1)} \sum_k \frac{g_1 \sqrt{L} \psi_k (z_1) \psi_k (z)}{\omega_b - \varepsilon_{k}} \\
    &= \sqrt{Z (\omega_b) L} g_1 G^{(0)} (z,z_1; \omega_b),
\end{align}
with $\alpha^{(1)} = \sqrt{Z (\omega_b)}$.

Analogously we find a bound state emerging due to the coupling of the waveguide only with the second qubit:
\begin{align}
    |\psi_{BS}^{(2)} \rangle = \alpha^{(2)} \left[ |ge\rangle |\text{vac}\rangle + |gg\rangle \sum_k \frac{g_2 \sqrt{L} \psi_k (z_2)}{\omega_b - \varepsilon_{k}} a_k^{\dagger}|\text{vac}\rangle \right].
    \label{BS2}
\end{align}
In the symmetric setup with $g_1 = g_2$, $\omega_{q,1} = \omega_{q,2}$, and $z_1 = - z_2$, the energy of this bound state has the same value $\omega_b$ as for the state \eqref{BS1}. 

When the both qubits are coupled to the waveguide, we can find the energy splitting of the two bound states from the overlap of the photonic contributions to the states \eqref{BS1} and \eqref{BS2}. In addition, we must subtract the matrix element of $H_0$, given in \eqref{H_0}, since this Hamiltonian has been already used twice, that is in the eigenvalue problems for each bound state. Thereby we get
\begin{align}
   &\langle \psi_{BS}^{(1)}| (\omega_b - H_0) |\psi_{BS}^{(2)} \rangle =  \langle \psi_{phot}^{(1)}| (\omega_b - H_0) |\psi_{phot}^{(2)} \rangle \\
   &= Z (\omega_b) g_1 g_2 L \sum_k \frac{\psi_k (z_1) \psi_k (z_2)}{\omega_b - \varepsilon_k} \equiv Z (\omega_b) \hat{\Sigma}_{12}^{(0)} (\omega_b). \nonumber
\end{align}

\section{Lattice realization of the waveguide}
\label{app:lattG}

\begin{figure}[t]
  \centering
  \includegraphics[scale=0.6]{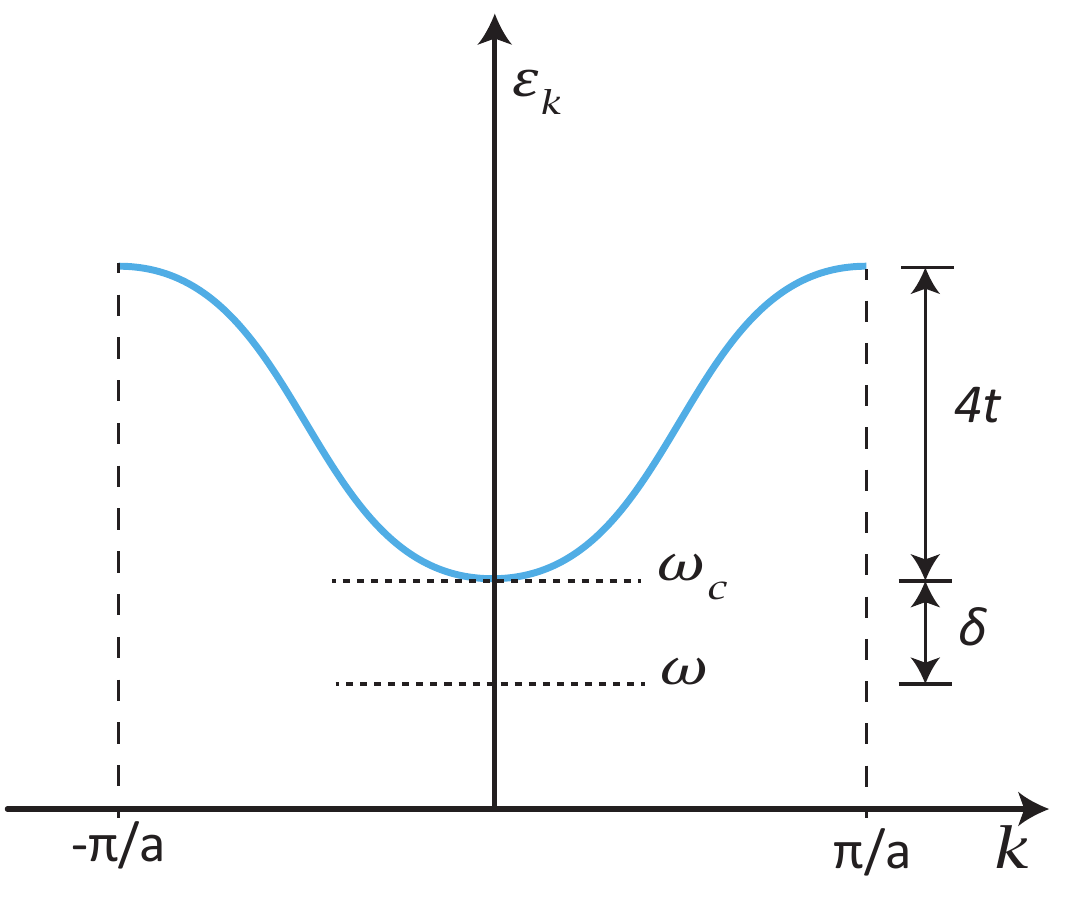}
  \caption{Dispersion relation of a waveguide modelled as cavity array. The the detuning is given by $\delta$ and the size of the passband is $2t$.}
  \label{dispersion_PhC}
 \end{figure}

Let us model a cavity array or a photonic crystal by a chain of $N$ sites (labelled by $n=1, \ldots,N$). Treating it in the tight-binding approximation we introduce the nearest-neighbor hopping amplitude $-t$ and the uniform on-site energy $\omega_c+2 t$.  The energy spectrum (see Fig.~(\ref{dispersion_PhC})) and the eigenfunctions of this model are
\begin{align}
    \varepsilon_k &= \omega_c +2t -2t \cos k a ,
    \label{eigenspectrum1} \\
    \psi_k (z_n) &= \sqrt{\frac{2}{L}} \sin (k z_n ),
    \label{eigenmodes1}
\end{align}
where $k=\frac{l \pi}{L}$ is labelled by integer $l$, $1 \leq l \leq N$. Hereby we introduced the chain's length $L = (N+1) a$ in terms of the lattice constant $a$, as well as the $n$th site coordinate $z_n = na$. We note the normalization $a \cdot \sum_{n=1}^N \psi_k^2 (z_n)=1$.

All observables discussed in the paper are expressed via the core object --- the waveguide's bare Green's function \eqref{G0_z1}. Its lattice analogue reads
\begin{align}
    & G^{(0)} (z_n , z_{n'}; \omega) \equiv G_{nn'}^{(0)} (\omega) \\
    = & \frac{\omega}{\omega_c} \sum_{l=1}^N \frac{\psi_{\frac{l \pi}{L}} (n a) \psi_{\frac{l \pi}{L}} (n' a)}{\omega - \varepsilon_{\frac{l \pi}{L}}} - \frac{\delta_{nn'}}{\omega_c a} \\
    \approx & \sum_{l=1}^N \frac{\psi_{\frac{l \pi}{L}} (n a) \psi_{\frac{l \pi}{L}} (n' a)}{\omega - \varepsilon_{\frac{l \pi}{L}}}.
\end{align}
This finite sum can be easily evaluated numerically.

For large $N$, we can replace the sum by the integral
\begin{align}
    G_{nn'}^{(0)} (\omega) & \approx \frac{2}{\pi} \int_0^{\pi/a} dk  \frac{\sin (k n a) \sin (k n' a)}{\omega - \omega_c -2t + 2t \cos k a}
    \label{int_approx} \\
    &= \frac{1}{a} \int_{-\pi}^{\pi} \frac{d \bar{k}}{2 \pi}  \frac{e^{i \bar{k} |n-n'|}-e^{i\bar{k} |n+n'|}}{\omega - \omega_c -2t + 2t \cos \bar{k}} \\
    & \equiv I_{|n-n'|} - I_{|n+n'|}.
\end{align}

Evaluating the integral $I_{\bar{n}}$ at integer $\bar{n} \geq 0$ and $\omega_c -\omega \equiv \delta >0$ (in practice, the condition should be $\delta \gg t a^2 \frac{\pi^2}{L^2}$ in order to justify the integral approximation \eqref{int_approx}), we obtain
\begin{align}
    I_{\bar{n}} &\equiv  \frac{1}{a} \int_{-\pi}^{\pi} \frac{d \bar{k}}{2 \pi}  \frac{e^{i \bar{k} \, \bar{n}}}{\omega - \omega_c -2t + 2t \cos \bar{k}} \\
    &=- \frac{1}{a\sqrt{(4 t + \delta) \delta}} \left( \frac{2t}{2t + \delta + \sqrt{ (4t + \delta) \delta}} \right)^{\bar{n}}.
\end{align}
For the broad bandwidth $4 t \gg \delta$ we further approximate
\begin{align}
    I_{\bar{n}} &\approx - \frac{1}{a \sqrt{4 t \delta}} e^{- \bar{n} \sqrt{\frac{\delta}{t}}} ,
\end{align}
and then
\begin{align}
    G_{nn'}^{(0)} (\omega) \approx I_{|n-n'|} \approx - \frac{1}{a \sqrt{4 t \delta}} e^{- |n-n'| \sqrt{\frac{\delta}{t}}}.
    \label{G0_phc}
\end{align}

In general, the low-energy approximation $\varepsilon_k \approx \omega_c + \frac{k^2}{2 m_{\text{eff}}}$ in terms of the effective mass $m_{\text{eff}}= \frac{1}{2 ta^2}$ directly leads to 
\eqref{G0_phc}. This formula is obtained from the integral
\begin{align}
    G_{nn'}^{(0)} (\omega) & \approx -  \int_{-\infty}^{\infty} \frac{d k}{2 \pi}  \frac{e^{i k a |n-n'|}}{\delta  + \frac{k^2}{2 m_{\text{eff}}}}
\end{align}
after closing the integration contour in the upper half-plane of complex $k$ and evaluating the residue value at the pole $k_p = i \sqrt{2 m_{\text{eff}}\delta}$.

To draw an analogy of the present tight-binding model with the 3D waveguide model \eqref{rect_wg_disp}, we note the value $m_{\text{eff}} = \frac{\omega_c}{c^2}$ for the latter case. Thereby we also recover \eqref{G0_long} at $\omega_c \gg \delta >0$, which coincides with \eqref{G0_phc} (up to the relabelling of the parameters). This observation implies that the bound state properties near the cutoff frequency $\omega_c$ in both the 3D wave\-guide model \eqref{rect_wg_disp} and the cavity array model \eqref{eigenspectrum1} are physically equivalent.

\section{Bound state as a two-level system}
\label{app:dynamics}

For a single qubit detuned deep in the stopband, we derive an effective description in terms of the two-level system coupled to the ports by eliminating the waveguide modes.

From \eqref{a_om1} it follows
\begin{align}
     &  \tilde{a}_k (\omega) =  \sqrt{L} \sum_{j=1,2} \sum_{k'} G^{(p)}_{kk'} (\omega) f_{k'} \psi_{k'} (z_j) g_{j} \tilde{\sigma}_-^{(j)} (\omega)  \nonumber \\
& + \sqrt{\frac{\Gamma_L}{\pi}} \sqrt{L} \sum_{k'} G^{(p)}_{kk'} (\omega) f_{k'} \psi_{k'} (z_L) \tilde{c}_{in} (\omega) .
\end{align}

Deep in the stopband we make the Markov approximation $G^{(p)}_{kk'} (\omega) \approx G^{(p)}_{kk'} (\omega_{q,1})$. After this we transform $\tilde{a}_k (\omega)$ back to the time domain, and insert $a_k (t)$ into \eqref{eq_Qa}, relaxing the weak-intensity approximation $\sigma_z^{(j)} \approx -1$.

For  a single qubit $j=1$ this gives
\begin{align}
    \frac{d}{dt} \sigma_-^{(1)} (t) &\approx  -i \left( \omega_{q,1} -i \frac{\Gamma_{q,1}}{2} - \hat{\Sigma}_{11} (\omega_{q,1}) \right) \sigma_-^{(1)} (t) \\
&+ i  g_{1}  \sigma_z^{(1)} (t)  \sqrt{\frac{\Gamma_L}{\pi}} L  G^{(p)} (z_1, z_L ;\omega_{q,1}) c_{in} (t).
\end{align}
Averaging over an initial coherent state and denoting 
\begin{align}
     2 g_{1}  \sqrt{\frac{\Gamma_L}{\pi}} L  G^{(p)} (z_1, z_L ;\omega_{q,1}) \langle c_{in} (t) \rangle = \Omega_r e^{-i \omega_d t}
\end{align}
in terms of the Rabi frequency $\Omega_r$ and the driving frequency $\omega_d$,
we obtain the equation
\begin{align}
     \frac{d}{dt} \langle \sigma_-^{(1)} (t)\rangle  &\approx  -i \left( \omega_{b} -i \frac{\Gamma_b}{2} \right) \langle \sigma_-^{(1)} (t) \rangle \\
&+ \frac{i}{2}  \Omega_r e^{-i \omega_d t}  \langle \sigma_z^{(1)} (t)  \rangle.
\end{align}
In addition, we derive
\begin{align}
     \frac{d}{dt} \langle \sigma_z^{(1)} (t)\rangle  &\approx  - \Gamma_b [ 1 + \langle \sigma_z^{(1)} (t) \rangle ] \\
&- i  \Omega_r   \langle e^{-i \omega_d t} \sigma_+^{(1)} (t)  - e^{i \omega_d t} \sigma_-^{(1)} (t) \rangle .
\end{align}

Inserting the approximate $a_k (t)$ into \eqref{out_a_exact} we obtain the effective input-output relation
\begin{align}
c_{out} (t)
&= - 2 i     \sqrt{\Gamma_R \Gamma_L} L  G^{(p)} (z_R, z_L ; \omega_b )  c_{in} (t)   \label{in_out_eff1.1}\\
& - 2 i   g_1 \sqrt{\pi \Gamma_R}  L   G^{(p)} (z_R , z_1 ; \omega_{b})  \sigma_-^{(1)} (t).
\label{in_out_eff1.2}
\end{align}

In the stopband the value of $ G^{(p)} (z_R, z_L ; \omega_b )$ is negligible, therefore $\langle  \sigma_-^{(1)} (t) \rangle$ is a direct measure of $\langle c_{out} (t) \rangle$. In the co-rotating frame, where $\langle  \hat{\sigma}_-^{(1)} (t) \rangle = \langle  \sigma_-^{(1)} (t) \rangle e^{i \omega_d t}$, $\langle  \hat{\sigma}_z^{(1)} (t) \rangle = \langle  \sigma_-^{(1)} (t) \rangle$, and $\langle \hat{c}_{out} (t) \rangle = \langle c_{out} (t) \rangle e^{i \omega_d t}$,  we relate
\begin{align}
\langle \hat{c}_{out} (t) \rangle
\approx - 2 i   g_1 \sqrt{\pi \Gamma_R}  L   G^{(p)} (z_R , z_1 ; \omega_{b})  \langle \hat{\sigma}_-^{(1)} (t) \rangle.
\label{in_out_eff2}
\end{align}
As well we get the following equations for the qubit observables
\begin{align}
    \frac{d}{dt} \langle \hat{\sigma}_-^{(1)} (t)\rangle  &\approx  i \left( \delta_d + i \frac{\Gamma_b}{2} \right) \langle \hat{\sigma}_-^{(1)} (t) \rangle + \frac{i}{2}  \Omega_r \langle \hat{\sigma}_z^{(1)} (t)  \rangle, \label{app:master_eq1} \\
     \frac{d}{dt} \langle \hat{\sigma}_z^{(1)} (t)\rangle  &\approx  - \Gamma_b [ 1 + \langle \hat{\sigma}_z^{(1)} (t) \rangle ] - i  \Omega_r   \langle \hat{\sigma}_+^{(1)} (t)  - \hat{\sigma}_-^{(1)} (t) \rangle ,
     \label{app:master_eq2}
\end{align} 
where $\delta_d = \omega_d - \omega_b$. On their basis we find the steady state value
\begin{align}
    |\langle \hat{\sigma}_-^{(1)} \rangle_{ss} | = |\langle \sigma_-^{(1)} \rangle_{ss} | = \frac{\Omega_r \sqrt{4 \delta_d^2 + \Gamma_b^2}}{2 \Omega_r^2 + 4 \delta_d^2 + \Gamma_b^2}.
    \label{app:ss_dyanmics}
\end{align}

\end{appendix}
\bibliographystyle{apsrev4-2}
\bibliography{apbs_rwg}

\begin{thebibliography}{39}%
\makeatletter
\providecommand \@ifxundefined [1]{%
 \@ifx{#1\undefined}
}%
\providecommand \@ifnum [1]{%
 \ifnum #1\expandafter \@firstoftwo
 \else \expandafter \@secondoftwo
 \fi
}%
\providecommand \@ifx [1]{%
 \ifx #1\expandafter \@firstoftwo
 \else \expandafter \@secondoftwo
 \fi
}%
\providecommand \natexlab [1]{#1}%
\providecommand \enquote  [1]{``#1''}%
\providecommand \bibnamefont  [1]{#1}%
\providecommand \bibfnamefont [1]{#1}%
\providecommand \citenamefont [1]{#1}%
\providecommand \href@noop [0]{\@secondoftwo}%
\providecommand \href [0]{\begingroup \@sanitize@url \@href}%
\providecommand \@href[1]{\@@startlink{#1}\@@href}%
\providecommand \@@href[1]{\endgroup#1\@@endlink}%
\providecommand \@sanitize@url [0]{\catcode `\\12\catcode `\$12\catcode
  `\&12\catcode `\#12\catcode `\^12\catcode `\_12\catcode `\%12\relax}%
\providecommand \@@startlink[1]{}%
\providecommand \@@endlink[0]{}%
\providecommand \url  [0]{\begingroup\@sanitize@url \@url }%
\providecommand \@url [1]{\endgroup\@href {#1}{\urlprefix }}%
\providecommand \urlprefix  [0]{URL }%
\providecommand \Eprint [0]{\href }%
\providecommand \doibase [0]{https://doi.org/}%
\providecommand \selectlanguage [0]{\@gobble}%
\providecommand \bibinfo  [0]{\@secondoftwo}%
\providecommand \bibfield  [0]{\@secondoftwo}%
\providecommand \translation [1]{[#1]}%
\providecommand \BibitemOpen [0]{}%
\providecommand \bibitemStop [0]{}%
\providecommand \bibitemNoStop [0]{.\EOS\space}%
\providecommand \EOS [0]{\spacefactor3000\relax}%
\providecommand \BibitemShut  [1]{\csname bibitem#1\endcsname}%
\let\auto@bib@innerbib\@empty
%</preamble>
\bibitem [{\citenamefont {Wallraff}\ \emph {et~al.}(2004)\citenamefont
  {Wallraff}, \citenamefont {Schuster}, \citenamefont {Blais}, \citenamefont
  {Frunzio}, \citenamefont {Huang}, \citenamefont {Majer}, \citenamefont
  {Kumar}, \citenamefont {Girvin},\ and\ \citenamefont
  {Schoelkopf}}]{Wallraff2004}%
  \BibitemOpen
  \bibfield  {author} {\bibinfo {author} {\bibfnamefont {A.}~\bibnamefont
  {Wallraff}}, \bibinfo {author} {\bibfnamefont {D.~I.}\ \bibnamefont
  {Schuster}}, \bibinfo {author} {\bibfnamefont {A.}~\bibnamefont {Blais}},
  \bibinfo {author} {\bibfnamefont {L.}~\bibnamefont {Frunzio}}, \bibinfo
  {author} {\bibfnamefont {R.-S.}\ \bibnamefont {Huang}}, \bibinfo {author}
  {\bibfnamefont {J.}~\bibnamefont {Majer}}, \bibinfo {author} {\bibfnamefont
  {S.}~\bibnamefont {Kumar}}, \bibinfo {author} {\bibfnamefont {S.~M.}\
  \bibnamefont {Girvin}},\ and\ \bibinfo {author} {\bibfnamefont {R.~J.}\
  \bibnamefont {Schoelkopf}},\ }\href {https://doi.org/10.1038/nature02851}
  {\bibfield  {journal} {\bibinfo  {journal} {Nature}\ }\textbf {\bibinfo
  {volume} {431}},\ \bibinfo {pages} {162} (\bibinfo {year}
  {2004})}\BibitemShut {NoStop}%
\bibitem [{\citenamefont {Arute}\ \emph {et~al.}(2019)\citenamefont {Arute},
  \citenamefont {Arya}, \citenamefont {Babbush}, \citenamefont {Bacon},
  \citenamefont {Bardin}, \citenamefont {Barends}, \citenamefont {Biswas},
  \citenamefont {Boixo}, \citenamefont {Brandao}, \citenamefont {Buell} \emph
  {et~al.}}]{arute2019}%
  \BibitemOpen
  \bibfield  {author} {\bibinfo {author} {\bibfnamefont {F.}~\bibnamefont
  {Arute}}, \bibinfo {author} {\bibfnamefont {K.}~\bibnamefont {Arya}},
  \bibinfo {author} {\bibfnamefont {R.}~\bibnamefont {Babbush}}, \bibinfo
  {author} {\bibfnamefont {D.}~\bibnamefont {Bacon}}, \bibinfo {author}
  {\bibfnamefont {J.~C.}\ \bibnamefont {Bardin}}, \bibinfo {author}
  {\bibfnamefont {R.}~\bibnamefont {Barends}}, \bibinfo {author} {\bibfnamefont
  {R.}~\bibnamefont {Biswas}}, \bibinfo {author} {\bibfnamefont
  {S.}~\bibnamefont {Boixo}}, \bibinfo {author} {\bibfnamefont {F.~G.}\
  \bibnamefont {Brandao}}, \bibinfo {author} {\bibfnamefont {D.~A.}\
  \bibnamefont {Buell}}, \emph {et~al.},\ }\href
  {https://doi.org/10.1038/s41586-019-1666-5} {\bibfield  {journal} {\bibinfo
  {journal} {Nature}\ }\textbf {\bibinfo {volume} {574}},\ \bibinfo {pages}
  {505} (\bibinfo {year} {2019})}\BibitemShut {NoStop}%
\bibitem [{\citenamefont {Sundaresan}\ \emph {et~al.}(2015)\citenamefont
  {Sundaresan}, \citenamefont {Liu}, \citenamefont {Sadri}, \citenamefont
  {Sz{\H{o}}cs}, \citenamefont {Underwood}, \citenamefont {Malekakhlagh},
  \citenamefont {T\"{u}reci},\ and\ \citenamefont {Houck}}]{sundaresan2015}%
  \BibitemOpen
  \bibfield  {author} {\bibinfo {author} {\bibfnamefont {N.~M.}\ \bibnamefont
  {Sundaresan}}, \bibinfo {author} {\bibfnamefont {Y.}~\bibnamefont {Liu}},
  \bibinfo {author} {\bibfnamefont {D.}~\bibnamefont {Sadri}}, \bibinfo
  {author} {\bibfnamefont {L.~J.}\ \bibnamefont {Sz{\H{o}}cs}}, \bibinfo
  {author} {\bibfnamefont {D.~L.}\ \bibnamefont {Underwood}}, \bibinfo {author}
  {\bibfnamefont {M.}~\bibnamefont {Malekakhlagh}}, \bibinfo {author}
  {\bibfnamefont {H.~E.}\ \bibnamefont {T\"{u}reci}},\ and\ \bibinfo {author}
  {\bibfnamefont {A.~A.}\ \bibnamefont {Houck}},\ }\href
  {https://doi.org/10.1103/physrevx.5.021035} {\bibfield  {journal} {\bibinfo
  {journal} {Physical Review X}\ }\textbf {\bibinfo {volume} {5}},\ \bibinfo
  {pages} {021035} (\bibinfo {year} {2015})}\BibitemShut {NoStop}%
\bibitem [{\citenamefont {Chakram}\ \emph {et~al.}(2021)\citenamefont
  {Chakram}, \citenamefont {Oriani}, \citenamefont {Naik}, \citenamefont
  {Dixit}, \citenamefont {He}, \citenamefont {Agrawal}, \citenamefont {Kwon},\
  and\ \citenamefont {Schuster}}]{Chakram2021}%
  \BibitemOpen
  \bibfield  {author} {\bibinfo {author} {\bibfnamefont {S.}~\bibnamefont
  {Chakram}}, \bibinfo {author} {\bibfnamefont {A.~E.}\ \bibnamefont {Oriani}},
  \bibinfo {author} {\bibfnamefont {R.~K.}\ \bibnamefont {Naik}}, \bibinfo
  {author} {\bibfnamefont {A.~V.}\ \bibnamefont {Dixit}}, \bibinfo {author}
  {\bibfnamefont {K.}~\bibnamefont {He}}, \bibinfo {author} {\bibfnamefont
  {A.}~\bibnamefont {Agrawal}}, \bibinfo {author} {\bibfnamefont
  {H.}~\bibnamefont {Kwon}},\ and\ \bibinfo {author} {\bibfnamefont {D.~I.}\
  \bibnamefont {Schuster}},\ }\href
  {https://doi.org/10.1103/physrevlett.127.107701} {\bibfield  {journal}
  {\bibinfo  {journal} {Physical Review Letters}\ }\textbf {\bibinfo {volume}
  {127}},\ \bibinfo {pages} {107701} (\bibinfo {year} {2021})}\BibitemShut
  {NoStop}%
\bibitem [{\citenamefont {Mart{\'{\i}}nez}\ \emph {et~al.}(2019)\citenamefont
  {Mart{\'{\i}}nez}, \citenamefont {L{\'{e}}ger}, \citenamefont {Gheeraert},
  \citenamefont {Dassonneville}, \citenamefont {Planat}, \citenamefont
  {Foroughi}, \citenamefont {Krupko}, \citenamefont {Buisson}, \citenamefont
  {Naud}, \citenamefont {Hasch-Guichard}, \citenamefont {Florens},
  \citenamefont {Snyman},\ and\ \citenamefont {Roch}}]{PuertasMartnez2019}%
  \BibitemOpen
  \bibfield  {author} {\bibinfo {author} {\bibfnamefont {J.~P.}\ \bibnamefont
  {Mart{\'{\i}}nez}}, \bibinfo {author} {\bibfnamefont {S.}~\bibnamefont
  {L{\'{e}}ger}}, \bibinfo {author} {\bibfnamefont {N.}~\bibnamefont
  {Gheeraert}}, \bibinfo {author} {\bibfnamefont {R.}~\bibnamefont
  {Dassonneville}}, \bibinfo {author} {\bibfnamefont {L.}~\bibnamefont
  {Planat}}, \bibinfo {author} {\bibfnamefont {F.}~\bibnamefont {Foroughi}},
  \bibinfo {author} {\bibfnamefont {Y.}~\bibnamefont {Krupko}}, \bibinfo
  {author} {\bibfnamefont {O.}~\bibnamefont {Buisson}}, \bibinfo {author}
  {\bibfnamefont {C.}~\bibnamefont {Naud}}, \bibinfo {author} {\bibfnamefont
  {W.}~\bibnamefont {Hasch-Guichard}}, \bibinfo {author} {\bibfnamefont
  {S.}~\bibnamefont {Florens}}, \bibinfo {author} {\bibfnamefont
  {I.}~\bibnamefont {Snyman}},\ and\ \bibinfo {author} {\bibfnamefont
  {N.}~\bibnamefont {Roch}},\ }\href
  {https://doi.org/10.1038/s41534-018-0104-0} {\bibfield  {journal} {\bibinfo
  {journal} {npj Quantum Information}\ }\textbf {\bibinfo {volume} {5}},\
  \bibinfo {pages} {1} (\bibinfo {year} {2019})}\BibitemShut {NoStop}%
\bibitem [{\citenamefont {Astafiev}\ \emph {et~al.}(2010)\citenamefont
  {Astafiev}, \citenamefont {Zagoskin}, \citenamefont {Abdumalikov},
  \citenamefont {Pashkin}, \citenamefont {Yamamoto}, \citenamefont {Inomata},
  \citenamefont {Nakamura},\ and\ \citenamefont {Tsai}}]{Astafiev2010}%
  \BibitemOpen
  \bibfield  {author} {\bibinfo {author} {\bibfnamefont {O.}~\bibnamefont
  {Astafiev}}, \bibinfo {author} {\bibfnamefont {A.~M.}\ \bibnamefont
  {Zagoskin}}, \bibinfo {author} {\bibfnamefont {A.~A.}\ \bibnamefont
  {Abdumalikov}}, \bibinfo {author} {\bibfnamefont {Y.~A.}\ \bibnamefont
  {Pashkin}}, \bibinfo {author} {\bibfnamefont {T.}~\bibnamefont {Yamamoto}},
  \bibinfo {author} {\bibfnamefont {K.}~\bibnamefont {Inomata}}, \bibinfo
  {author} {\bibfnamefont {Y.}~\bibnamefont {Nakamura}},\ and\ \bibinfo
  {author} {\bibfnamefont {J.~S.}\ \bibnamefont {Tsai}},\ }\href
  {https://doi.org/10.1126/science.1181918} {\bibfield  {journal} {\bibinfo
  {journal} {Science}\ }\textbf {\bibinfo {volume} {327}},\ \bibinfo {pages}
  {840} (\bibinfo {year} {2010})}\BibitemShut {NoStop}%
\bibitem [{\citenamefont {Abdumalikov}\ \emph {et~al.}(2011)\citenamefont
  {Abdumalikov}, \citenamefont {Astafiev}, \citenamefont {Pashkin},
  \citenamefont {Nakamura},\ and\ \citenamefont {Tsai}}]{abdumalikov2011}%
  \BibitemOpen
  \bibfield  {author} {\bibinfo {author} {\bibfnamefont {A.~A.}\ \bibnamefont
  {Abdumalikov}}, \bibinfo {author} {\bibfnamefont {O.~V.}\ \bibnamefont
  {Astafiev}}, \bibinfo {author} {\bibfnamefont {Y.~A.}\ \bibnamefont
  {Pashkin}}, \bibinfo {author} {\bibfnamefont {Y.}~\bibnamefont {Nakamura}},\
  and\ \bibinfo {author} {\bibfnamefont {J.~S.}\ \bibnamefont {Tsai}},\ }\href
  {https://doi.org/10.1103/physrevlett.107.043604} {\bibfield  {journal}
  {\bibinfo  {journal} {Physical Review Letters}\ }\textbf {\bibinfo {volume}
  {107}},\ \bibinfo {pages} {043604} (\bibinfo {year} {2011})}\BibitemShut
  {NoStop}%
\bibitem [{\citenamefont {Van~Loo}\ \emph {et~al.}(2013)\citenamefont
  {Van~Loo}, \citenamefont {Fedorov}, \citenamefont {Lalumiere}, \citenamefont
  {Sanders}, \citenamefont {Blais},\ and\ \citenamefont
  {Wallraff}}]{van2013photon}%
  \BibitemOpen
  \bibfield  {author} {\bibinfo {author} {\bibfnamefont {A.~F.}\ \bibnamefont
  {Van~Loo}}, \bibinfo {author} {\bibfnamefont {A.}~\bibnamefont {Fedorov}},
  \bibinfo {author} {\bibfnamefont {K.}~\bibnamefont {Lalumiere}}, \bibinfo
  {author} {\bibfnamefont {B.~C.}\ \bibnamefont {Sanders}}, \bibinfo {author}
  {\bibfnamefont {A.}~\bibnamefont {Blais}},\ and\ \bibinfo {author}
  {\bibfnamefont {A.}~\bibnamefont {Wallraff}},\ }\href@noop {} {\bibfield
  {journal} {\bibinfo  {journal} {Science}\ }\textbf {\bibinfo {volume}
  {342}},\ \bibinfo {pages} {1494} (\bibinfo {year} {2013})}\BibitemShut
  {NoStop}%
\bibitem [{\citenamefont {Lalumi{\`{e}}re}\ \emph {et~al.}(2013)\citenamefont
  {Lalumi{\`{e}}re}, \citenamefont {Sanders}, \citenamefont {van Loo},
  \citenamefont {Fedorov}, \citenamefont {Wallraff},\ and\ \citenamefont
  {Blais}}]{Lalumire2013}%
  \BibitemOpen
  \bibfield  {author} {\bibinfo {author} {\bibfnamefont {K.}~\bibnamefont
  {Lalumi{\`{e}}re}}, \bibinfo {author} {\bibfnamefont {B.~C.}\ \bibnamefont
  {Sanders}}, \bibinfo {author} {\bibfnamefont {A.~F.}\ \bibnamefont {van
  Loo}}, \bibinfo {author} {\bibfnamefont {A.}~\bibnamefont {Fedorov}},
  \bibinfo {author} {\bibfnamefont {A.}~\bibnamefont {Wallraff}},\ and\
  \bibinfo {author} {\bibfnamefont {A.}~\bibnamefont {Blais}},\ }\href
  {https://doi.org/10.1103/physreva.88.043806} {\bibfield  {journal} {\bibinfo
  {journal} {Physical Review A}\ }\textbf {\bibinfo {volume} {88}},\ \bibinfo
  {pages} {043806} (\bibinfo {year} {2013})}\BibitemShut {NoStop}%
\bibitem [{\citenamefont {Kimble}(2008)}]{Kimble2008}%
  \BibitemOpen
  \bibfield  {author} {\bibinfo {author} {\bibfnamefont {H.~J.}\ \bibnamefont
  {Kimble}},\ }\href {https://doi.org/10.1038/nature07127} {\bibfield
  {journal} {\bibinfo  {journal} {Nature}\ }\textbf {\bibinfo {volume} {453}},\
  \bibinfo {pages} {1023} (\bibinfo {year} {2008})}\BibitemShut {NoStop}%
\bibitem [{\citenamefont {Sipahigil}\ \emph {et~al.}(2016)\citenamefont
  {Sipahigil}, \citenamefont {Evans}, \citenamefont {Sukachev}, \citenamefont
  {Burek}, \citenamefont {Borregaard}, \citenamefont {Bhaskar}, \citenamefont
  {Nguyen}, \citenamefont {Pacheco}, \citenamefont {Atikian}, \citenamefont
  {Meuwly}, \citenamefont {Camacho}, \citenamefont {Jelezko}, \citenamefont
  {Bielejec}, \citenamefont {Park}, \citenamefont {Lon{\v{c}}ar},\ and\
  \citenamefont {Lukin}}]{Sipahigil2016}%
  \BibitemOpen
  \bibfield  {author} {\bibinfo {author} {\bibfnamefont {A.}~\bibnamefont
  {Sipahigil}}, \bibinfo {author} {\bibfnamefont {R.~E.}\ \bibnamefont
  {Evans}}, \bibinfo {author} {\bibfnamefont {D.~D.}\ \bibnamefont {Sukachev}},
  \bibinfo {author} {\bibfnamefont {M.~J.}\ \bibnamefont {Burek}}, \bibinfo
  {author} {\bibfnamefont {J.}~\bibnamefont {Borregaard}}, \bibinfo {author}
  {\bibfnamefont {M.~K.}\ \bibnamefont {Bhaskar}}, \bibinfo {author}
  {\bibfnamefont {C.~T.}\ \bibnamefont {Nguyen}}, \bibinfo {author}
  {\bibfnamefont {J.~L.}\ \bibnamefont {Pacheco}}, \bibinfo {author}
  {\bibfnamefont {H.~A.}\ \bibnamefont {Atikian}}, \bibinfo {author}
  {\bibfnamefont {C.}~\bibnamefont {Meuwly}}, \bibinfo {author} {\bibfnamefont
  {R.~M.}\ \bibnamefont {Camacho}}, \bibinfo {author} {\bibfnamefont
  {F.}~\bibnamefont {Jelezko}}, \bibinfo {author} {\bibfnamefont
  {E.}~\bibnamefont {Bielejec}}, \bibinfo {author} {\bibfnamefont
  {H.}~\bibnamefont {Park}}, \bibinfo {author} {\bibfnamefont {M.}~\bibnamefont
  {Lon{\v{c}}ar}},\ and\ \bibinfo {author} {\bibfnamefont {M.~D.}\ \bibnamefont
  {Lukin}},\ }\href {https://doi.org/10.1126/science.aah6875} {\bibfield
  {journal} {\bibinfo  {journal} {Science}\ }\textbf {\bibinfo {volume}
  {354}},\ \bibinfo {pages} {847} (\bibinfo {year} {2016})}\BibitemShut
  {NoStop}%
\bibitem [{\citenamefont {Pichler}\ \emph {et~al.}(2015)\citenamefont
  {Pichler}, \citenamefont {Ramos}, \citenamefont {Daley},\ and\ \citenamefont
  {Zoller}}]{Pichler2015}%
  \BibitemOpen
  \bibfield  {author} {\bibinfo {author} {\bibfnamefont {H.}~\bibnamefont
  {Pichler}}, \bibinfo {author} {\bibfnamefont {T.}~\bibnamefont {Ramos}},
  \bibinfo {author} {\bibfnamefont {A.~J.}\ \bibnamefont {Daley}},\ and\
  \bibinfo {author} {\bibfnamefont {P.}~\bibnamefont {Zoller}},\ }\href
  {https://doi.org/10.1103/physreva.91.042116} {\bibfield  {journal} {\bibinfo
  {journal} {Physical Review A}\ }\textbf {\bibinfo {volume} {91}},\ \bibinfo
  {pages} {042116} (\bibinfo {year} {2015})}\BibitemShut {NoStop}%
\bibitem [{\citenamefont {Hamann}\ \emph {et~al.}(2018)\citenamefont {Hamann},
  \citenamefont {M\"{u}ller}, \citenamefont {Jerger}, \citenamefont {Zanner},
  \citenamefont {Combes}, \citenamefont {Pletyukhov}, \citenamefont {Weides},
  \citenamefont {Stace},\ and\ \citenamefont {Fedorov}}]{RosarioHamann2018}%
  \BibitemOpen
  \bibfield  {author} {\bibinfo {author} {\bibfnamefont {A.~R.}\ \bibnamefont
  {Hamann}}, \bibinfo {author} {\bibfnamefont {C.}~\bibnamefont {M\"{u}ller}},
  \bibinfo {author} {\bibfnamefont {M.}~\bibnamefont {Jerger}}, \bibinfo
  {author} {\bibfnamefont {M.}~\bibnamefont {Zanner}}, \bibinfo {author}
  {\bibfnamefont {J.}~\bibnamefont {Combes}}, \bibinfo {author} {\bibfnamefont
  {M.}~\bibnamefont {Pletyukhov}}, \bibinfo {author} {\bibfnamefont
  {M.}~\bibnamefont {Weides}}, \bibinfo {author} {\bibfnamefont {T.~M.}\
  \bibnamefont {Stace}},\ and\ \bibinfo {author} {\bibfnamefont
  {A.}~\bibnamefont {Fedorov}},\ }\href
  {https://doi.org/10.1103/physrevlett.121.123601} {\bibfield  {journal}
  {\bibinfo  {journal} {Physical Review Letters}\ }\textbf {\bibinfo {volume}
  {121}},\ \bibinfo {pages} {123601} (\bibinfo {year} {2018})}\BibitemShut
  {NoStop}%
\bibitem [{\citenamefont {Hoi}\ \emph {et~al.}(2011)\citenamefont {Hoi},
  \citenamefont {Wilson}, \citenamefont {Johansson}, \citenamefont {Palomaki},
  \citenamefont {Peropadre},\ and\ \citenamefont {Delsing}}]{Hoi2011}%
  \BibitemOpen
  \bibfield  {author} {\bibinfo {author} {\bibfnamefont {I.-C.}\ \bibnamefont
  {Hoi}}, \bibinfo {author} {\bibfnamefont {C.~M.}\ \bibnamefont {Wilson}},
  \bibinfo {author} {\bibfnamefont {G.}~\bibnamefont {Johansson}}, \bibinfo
  {author} {\bibfnamefont {T.}~\bibnamefont {Palomaki}}, \bibinfo {author}
  {\bibfnamefont {B.}~\bibnamefont {Peropadre}},\ and\ \bibinfo {author}
  {\bibfnamefont {P.}~\bibnamefont {Delsing}},\ }\href
  {https://doi.org/10.1103/physrevlett.107.073601} {\bibfield  {journal}
  {\bibinfo  {journal} {Physical Review Letters}\ }\textbf {\bibinfo {volume}
  {107}},\ \bibinfo {pages} {073601} (\bibinfo {year} {2011})}\BibitemShut
  {NoStop}%
\bibitem [{\citenamefont {Bykov}(1975)}]{bykov1975}%
  \BibitemOpen
  \bibfield  {author} {\bibinfo {author} {\bibfnamefont {V.~P.}\ \bibnamefont
  {Bykov}},\ }\href
  {https://iopscience.iop.org/article/10.1070/QE1975v004n07ABEH009654}
  {\bibfield  {journal} {\bibinfo  {journal} {Soviet Journal of Quantum
  Electronics}\ }\textbf {\bibinfo {volume} {4}},\ \bibinfo {pages} {861}
  (\bibinfo {year} {1975})}\BibitemShut {NoStop}%
\bibitem [{\citenamefont {John}\ and\ \citenamefont {Wang}(1990)}]{John1990}%
  \BibitemOpen
  \bibfield  {author} {\bibinfo {author} {\bibfnamefont {S.}~\bibnamefont
  {John}}\ and\ \bibinfo {author} {\bibfnamefont {J.}~\bibnamefont {Wang}},\
  }\href {https://doi.org/10.1103/physrevlett.64.2418} {\bibfield  {journal}
  {\bibinfo  {journal} {Physical Review Letters}\ }\textbf {\bibinfo {volume}
  {64}},\ \bibinfo {pages} {2418} (\bibinfo {year} {1990})}\BibitemShut
  {NoStop}%
\bibitem [{\citenamefont {Kofman}\ \emph {et~al.}(1994)\citenamefont {Kofman},
  \citenamefont {Kurizki},\ and\ \citenamefont {Sherman}}]{Kofman1994}%
  \BibitemOpen
  \bibfield  {author} {\bibinfo {author} {\bibfnamefont {A.}~\bibnamefont
  {Kofman}}, \bibinfo {author} {\bibfnamefont {G.}~\bibnamefont {Kurizki}},\
  and\ \bibinfo {author} {\bibfnamefont {B.}~\bibnamefont {Sherman}},\ }\href
  {https://doi.org/10.1080/09500349414550381} {\bibfield  {journal} {\bibinfo
  {journal} {Journal of Modern Optics}\ }\textbf {\bibinfo {volume} {41}},\
  \bibinfo {pages} {353} (\bibinfo {year} {1994})}\BibitemShut {NoStop}%
\bibitem [{\citenamefont {Calaj{\'{o}}}\ \emph {et~al.}(2016)\citenamefont
  {Calaj{\'{o}}}, \citenamefont {Ciccarello}, \citenamefont {Chang},\ and\
  \citenamefont {Rabl}}]{calaj2016}%
  \BibitemOpen
  \bibfield  {author} {\bibinfo {author} {\bibfnamefont {G.}~\bibnamefont
  {Calaj{\'{o}}}}, \bibinfo {author} {\bibfnamefont {F.}~\bibnamefont
  {Ciccarello}}, \bibinfo {author} {\bibfnamefont {D.}~\bibnamefont {Chang}},\
  and\ \bibinfo {author} {\bibfnamefont {P.}~\bibnamefont {Rabl}},\ }\href
  {https://doi.org/10.1103/physreva.93.033833} {\bibfield  {journal} {\bibinfo
  {journal} {Physical Review A}\ }\textbf {\bibinfo {volume} {93}},\ \bibinfo
  {pages} {033833} (\bibinfo {year} {2016})}\BibitemShut {NoStop}%
\bibitem [{\citenamefont {Liu}\ and\ \citenamefont {Houck}(2016)}]{Liu2016}%
  \BibitemOpen
  \bibfield  {author} {\bibinfo {author} {\bibfnamefont {Y.}~\bibnamefont
  {Liu}}\ and\ \bibinfo {author} {\bibfnamefont {A.~A.}\ \bibnamefont
  {Houck}},\ }\href {https://doi.org/10.1038/nphys3834} {\bibfield  {journal}
  {\bibinfo  {journal} {Nature Physics}\ }\textbf {\bibinfo {volume} {13}},\
  \bibinfo {pages} {48} (\bibinfo {year} {2016})}\BibitemShut {NoStop}%
\bibitem [{\citenamefont {Sundaresan}\ \emph {et~al.}(2019)\citenamefont
  {Sundaresan}, \citenamefont {Lundgren}, \citenamefont {Zhu}, \citenamefont
  {Gorshkov},\ and\ \citenamefont {Houck}}]{sundaresan2019}%
  \BibitemOpen
  \bibfield  {author} {\bibinfo {author} {\bibfnamefont {N.~M.}\ \bibnamefont
  {Sundaresan}}, \bibinfo {author} {\bibfnamefont {R.}~\bibnamefont
  {Lundgren}}, \bibinfo {author} {\bibfnamefont {G.}~\bibnamefont {Zhu}},
  \bibinfo {author} {\bibfnamefont {A.~V.}\ \bibnamefont {Gorshkov}},\ and\
  \bibinfo {author} {\bibfnamefont {A.~A.}\ \bibnamefont {Houck}},\ }\href
  {https://doi.org/10.1103/physrevx.9.011021} {\bibfield  {journal} {\bibinfo
  {journal} {Physical Review X}\ }\textbf {\bibinfo {volume} {9}},\ \bibinfo
  {pages} {011021} (\bibinfo {year} {2019})}\BibitemShut {NoStop}%
\bibitem [{\citenamefont {Scigliuzzo}\ \emph {et~al.}(2021)\citenamefont
  {Scigliuzzo}, \citenamefont {Calaj{\`o}}, \citenamefont {Ciccarello},
  \citenamefont {Lozano}, \citenamefont {Bengtsson}, \citenamefont {Scarlino},
  \citenamefont {Wallraff}, \citenamefont {Chang}, \citenamefont {Delsing},\
  and\ \citenamefont {Gasparinetti}}]{scigliuzzo2021}%
  \BibitemOpen
  \bibfield  {author} {\bibinfo {author} {\bibfnamefont {M.}~\bibnamefont
  {Scigliuzzo}}, \bibinfo {author} {\bibfnamefont {G.}~\bibnamefont
  {Calaj{\`o}}}, \bibinfo {author} {\bibfnamefont {F.}~\bibnamefont
  {Ciccarello}}, \bibinfo {author} {\bibfnamefont {D.~P.}\ \bibnamefont
  {Lozano}}, \bibinfo {author} {\bibfnamefont {A.}~\bibnamefont {Bengtsson}},
  \bibinfo {author} {\bibfnamefont {P.}~\bibnamefont {Scarlino}}, \bibinfo
  {author} {\bibfnamefont {A.}~\bibnamefont {Wallraff}}, \bibinfo {author}
  {\bibfnamefont {D.}~\bibnamefont {Chang}}, \bibinfo {author} {\bibfnamefont
  {P.}~\bibnamefont {Delsing}},\ and\ \bibinfo {author} {\bibfnamefont
  {S.}~\bibnamefont {Gasparinetti}},\ }\href {https://arxiv.org/abs/2107.06852}
  {\bibfield  {journal} {\bibinfo  {journal} {arXiv:2107.06852}\ } (\bibinfo
  {year} {2021})}\BibitemShut {NoStop}%
\bibitem [{\citenamefont {Zhang}\ \emph {et~al.}(2022)\citenamefont {Zhang},
  \citenamefont {Kim}, \citenamefont {Mark}, \citenamefont {Choi},\ and\
  \citenamefont {Painter}}]{zhang2022}%
  \BibitemOpen
  \bibfield  {author} {\bibinfo {author} {\bibfnamefont {X.}~\bibnamefont
  {Zhang}}, \bibinfo {author} {\bibfnamefont {E.}~\bibnamefont {Kim}}, \bibinfo
  {author} {\bibfnamefont {D.~K.}\ \bibnamefont {Mark}}, \bibinfo {author}
  {\bibfnamefont {S.}~\bibnamefont {Choi}},\ and\ \bibinfo {author}
  {\bibfnamefont {O.}~\bibnamefont {Painter}},\ }\href
  {https://arxiv.org/abs/2206.12803} {\bibfield  {journal} {\bibinfo  {journal}
  {arXiv: 2206.12803}\ } (\bibinfo {year} {2022})}\BibitemShut {NoStop}%
\bibitem [{\citenamefont {John}\ and\ \citenamefont {Quang}(1996)}]{John1996}%
  \BibitemOpen
  \bibfield  {author} {\bibinfo {author} {\bibfnamefont {S.}~\bibnamefont
  {John}}\ and\ \bibinfo {author} {\bibfnamefont {T.}~\bibnamefont {Quang}},\
  }\href {https://doi.org/10.1103/physrevlett.76.1320} {\bibfield  {journal}
  {\bibinfo  {journal} {Physical Review Letters}\ }\textbf {\bibinfo {volume}
  {76}},\ \bibinfo {pages} {1320} (\bibinfo {year} {1996})}\BibitemShut
  {NoStop}%
\bibitem [{\citenamefont {Douglas}\ \emph {et~al.}(2015)\citenamefont
  {Douglas}, \citenamefont {Habibian}, \citenamefont {Hung}, \citenamefont
  {Gorshkov}, \citenamefont {Kimble},\ and\ \citenamefont
  {Chang}}]{Douglas2015}%
  \BibitemOpen
  \bibfield  {author} {\bibinfo {author} {\bibfnamefont {J.~S.}\ \bibnamefont
  {Douglas}}, \bibinfo {author} {\bibfnamefont {H.}~\bibnamefont {Habibian}},
  \bibinfo {author} {\bibfnamefont {C.-L.}\ \bibnamefont {Hung}}, \bibinfo
  {author} {\bibfnamefont {A.~V.}\ \bibnamefont {Gorshkov}}, \bibinfo {author}
  {\bibfnamefont {H.~J.}\ \bibnamefont {Kimble}},\ and\ \bibinfo {author}
  {\bibfnamefont {D.~E.}\ \bibnamefont {Chang}},\ }\href
  {https://doi.org/10.1038/nphoton.2015.57} {\bibfield  {journal} {\bibinfo
  {journal} {Nature Photonics}\ }\textbf {\bibinfo {volume} {9}},\ \bibinfo
  {pages} {326} (\bibinfo {year} {2015})}\BibitemShut {NoStop}%
\bibitem [{\citenamefont {Shi}\ \emph {et~al.}(2016)\citenamefont {Shi},
  \citenamefont {Wu}, \citenamefont {Gonz{\'{a}}lez-Tudela},\ and\
  \citenamefont {Cirac}}]{shi2016}%
  \BibitemOpen
  \bibfield  {author} {\bibinfo {author} {\bibfnamefont {T.}~\bibnamefont
  {Shi}}, \bibinfo {author} {\bibfnamefont {Y.-H.}\ \bibnamefont {Wu}},
  \bibinfo {author} {\bibfnamefont {A.}~\bibnamefont {Gonz{\'{a}}lez-Tudela}},\
  and\ \bibinfo {author} {\bibfnamefont {J.}~\bibnamefont {Cirac}},\ }\href
  {https://doi.org/10.1103/physrevx.6.021027} {\bibfield  {journal} {\bibinfo
  {journal} {Physical Review X}\ }\textbf {\bibinfo {volume} {6}},\ \bibinfo
  {pages} {021027} (\bibinfo {year} {2016})}\BibitemShut {NoStop}%
\bibitem [{\citenamefont {Shi}\ \emph {et~al.}(2018)\citenamefont {Shi},
  \citenamefont {Wu}, \citenamefont {Gonz{\'{a}}lez-Tudela},\ and\
  \citenamefont {Cirac}}]{Shi2018}%
  \BibitemOpen
  \bibfield  {author} {\bibinfo {author} {\bibfnamefont {T.}~\bibnamefont
  {Shi}}, \bibinfo {author} {\bibfnamefont {Y.-H.}\ \bibnamefont {Wu}},
  \bibinfo {author} {\bibfnamefont {A.}~\bibnamefont {Gonz{\'{a}}lez-Tudela}},\
  and\ \bibinfo {author} {\bibfnamefont {J.~I.}\ \bibnamefont {Cirac}},\ }\href
  {https://doi.org/10.1088/1367-2630/aae4a9} {\bibfield  {journal} {\bibinfo
  {journal} {New Journal of Physics}\ }\textbf {\bibinfo {volume} {20}},\
  \bibinfo {pages} {105005} (\bibinfo {year} {2018})}\BibitemShut {NoStop}%
\bibitem [{\citenamefont {Mirhosseini}\ \emph {et~al.}(2018)\citenamefont
  {Mirhosseini}, \citenamefont {Kim}, \citenamefont {Ferreira}, \citenamefont
  {Kalaee}, \citenamefont {Sipahigil}, \citenamefont {Keller},\ and\
  \citenamefont {Painter}}]{Mirhosseini2018}%
  \BibitemOpen
  \bibfield  {author} {\bibinfo {author} {\bibfnamefont {M.}~\bibnamefont
  {Mirhosseini}}, \bibinfo {author} {\bibfnamefont {E.}~\bibnamefont {Kim}},
  \bibinfo {author} {\bibfnamefont {V.~S.}\ \bibnamefont {Ferreira}}, \bibinfo
  {author} {\bibfnamefont {M.}~\bibnamefont {Kalaee}}, \bibinfo {author}
  {\bibfnamefont {A.}~\bibnamefont {Sipahigil}}, \bibinfo {author}
  {\bibfnamefont {A.~J.}\ \bibnamefont {Keller}},\ and\ \bibinfo {author}
  {\bibfnamefont {O.}~\bibnamefont {Painter}},\ }\href
  {https://doi.org/10.1038/s41467-018-06142-z} {\bibfield  {journal} {\bibinfo
  {journal} {Nature Communications}\ }\textbf {\bibinfo {volume} {9}},\
  \bibinfo {pages} {1} (\bibinfo {year} {2018})}\BibitemShut {NoStop}%
\bibitem [{\citenamefont {Pozar}(2011)}]{pozar2011microwave}%
  \BibitemOpen
  \bibfield  {author} {\bibinfo {author} {\bibfnamefont {D.~M.}\ \bibnamefont
  {Pozar}},\ }\href
  {https://www.wiley.com/en-au/Microwave+Engineering,+4th+Edition-p-9780470631553}
  {\emph {\bibinfo {title} {Microwave engineering}}}\ (\bibinfo  {publisher}
  {John wiley \& sons},\ \bibinfo {year} {2011})\BibitemShut {NoStop}%
\bibitem [{\citenamefont {Shahmoon}\ and\ \citenamefont
  {Kurizki}(2013)}]{Shahmoon2013}%
  \BibitemOpen
  \bibfield  {author} {\bibinfo {author} {\bibfnamefont {E.}~\bibnamefont
  {Shahmoon}}\ and\ \bibinfo {author} {\bibfnamefont {G.}~\bibnamefont
  {Kurizki}},\ }\href {https://doi.org/10.1103/physreva.87.033831} {\bibfield
  {journal} {\bibinfo  {journal} {Physical Review A}\ }\textbf {\bibinfo
  {volume} {87}},\ \bibinfo {pages} {033831} (\bibinfo {year}
  {2013})}\BibitemShut {NoStop}%
\bibitem [{\citenamefont {Paik}\ \emph {et~al.}(2011)\citenamefont {Paik},
  \citenamefont {Schuster}, \citenamefont {Bishop}, \citenamefont {Kirchmair},
  \citenamefont {Catelani}, \citenamefont {Sears}, \citenamefont {Johnson},
  \citenamefont {Reagor}, \citenamefont {Frunzio}, \citenamefont {Glazman},
  \citenamefont {Girvin}, \citenamefont {Devoret},\ and\ \citenamefont
  {Schoelkopf}}]{Paik2011}%
  \BibitemOpen
  \bibfield  {author} {\bibinfo {author} {\bibfnamefont {H.}~\bibnamefont
  {Paik}}, \bibinfo {author} {\bibfnamefont {D.~I.}\ \bibnamefont {Schuster}},
  \bibinfo {author} {\bibfnamefont {L.~S.}\ \bibnamefont {Bishop}}, \bibinfo
  {author} {\bibfnamefont {G.}~\bibnamefont {Kirchmair}}, \bibinfo {author}
  {\bibfnamefont {G.}~\bibnamefont {Catelani}}, \bibinfo {author}
  {\bibfnamefont {A.~P.}\ \bibnamefont {Sears}}, \bibinfo {author}
  {\bibfnamefont {B.~R.}\ \bibnamefont {Johnson}}, \bibinfo {author}
  {\bibfnamefont {M.~J.}\ \bibnamefont {Reagor}}, \bibinfo {author}
  {\bibfnamefont {L.}~\bibnamefont {Frunzio}}, \bibinfo {author} {\bibfnamefont
  {L.~I.}\ \bibnamefont {Glazman}}, \bibinfo {author} {\bibfnamefont {S.~M.}\
  \bibnamefont {Girvin}}, \bibinfo {author} {\bibfnamefont {M.~H.}\
  \bibnamefont {Devoret}},\ and\ \bibinfo {author} {\bibfnamefont {R.~J.}\
  \bibnamefont {Schoelkopf}},\ }\href
  {https://doi.org/10.1103/physrevlett.107.240501} {\bibfield  {journal}
  {\bibinfo  {journal} {Physical Review Letters}\ }\textbf {\bibinfo {volume}
  {107}},\ \bibinfo {pages} {240501} (\bibinfo {year} {2011})}\BibitemShut
  {NoStop}%
\bibitem [{\citenamefont {Asenjo-Garcia}\ \emph {et~al.}(2017)\citenamefont
  {Asenjo-Garcia}, \citenamefont {Hood}, \citenamefont {Chang},\ and\
  \citenamefont {Kimble}}]{AsenjoGarcia2017}%
  \BibitemOpen
  \bibfield  {author} {\bibinfo {author} {\bibfnamefont {A.}~\bibnamefont
  {Asenjo-Garcia}}, \bibinfo {author} {\bibfnamefont {J.~D.}\ \bibnamefont
  {Hood}}, \bibinfo {author} {\bibfnamefont {D.~E.}\ \bibnamefont {Chang}},\
  and\ \bibinfo {author} {\bibfnamefont {H.~J.}\ \bibnamefont {Kimble}},\
  }\href {https://doi.org/10.1103/physreva.95.033818} {\bibfield  {journal}
  {\bibinfo  {journal} {Physical Review A}\ }\textbf {\bibinfo {volume} {95}},\
  \bibinfo {pages} {033818} (\bibinfo {year} {2017})}\BibitemShut {NoStop}%
\bibitem [{\citenamefont {Schneider}\ \emph {et~al.}(2016)\citenamefont
  {Schneider}, \citenamefont {Sproll}, \citenamefont {Stawiarski},
  \citenamefont {Schmitteckert},\ and\ \citenamefont {Busch}}]{Schneider2016}%
  \BibitemOpen
  \bibfield  {author} {\bibinfo {author} {\bibfnamefont {M.~P.}\ \bibnamefont
  {Schneider}}, \bibinfo {author} {\bibfnamefont {T.}~\bibnamefont {Sproll}},
  \bibinfo {author} {\bibfnamefont {C.}~\bibnamefont {Stawiarski}}, \bibinfo
  {author} {\bibfnamefont {P.}~\bibnamefont {Schmitteckert}},\ and\ \bibinfo
  {author} {\bibfnamefont {K.}~\bibnamefont {Busch}},\ }\bibfield  {journal}
  {\bibinfo  {journal} {Physical Review A}\ }\textbf {\bibinfo {volume} {93}},\
  \href {https://doi.org/10.1103/physreva.93.013828}
  {10.1103/physreva.93.013828} (\bibinfo {year} {2016})\BibitemShut {NoStop}%
\bibitem [{\citenamefont {Gardiner}\ \emph {et~al.}(2004)\citenamefont
  {Gardiner}, \citenamefont {Zoller},\ and\ \citenamefont
  {Zoller}}]{gardiner2004quantum}%
  \BibitemOpen
  \bibfield  {author} {\bibinfo {author} {\bibfnamefont {C.}~\bibnamefont
  {Gardiner}}, \bibinfo {author} {\bibfnamefont {P.}~\bibnamefont {Zoller}},\
  and\ \bibinfo {author} {\bibfnamefont {P.}~\bibnamefont {Zoller}},\ }\href
  {https://link.springer.com/book/9783540223016} {\emph {\bibinfo {title}
  {Quantum noise}}}\ (\bibinfo  {publisher} {Springer Science \& Business
  Media},\ \bibinfo {year} {2004})\BibitemShut {NoStop}%
\bibitem [{\citenamefont {Mahan}(2000)}]{Mahan2000ManyParticle}%
  \BibitemOpen
  \bibfield  {author} {\bibinfo {author} {\bibfnamefont {D.~G.}\ \bibnamefont
  {Mahan}},\ }\href {https://link.springer.com/book/10.1007/978-1-4757-5714-9}
  {\emph {\bibinfo {title} {Many-Particle Physics}}}\ (\bibinfo  {publisher}
  {Springer New York, NY},\ \bibinfo {year} {2000})\BibitemShut {NoStop}%
\bibitem [{\citenamefont {Houck}\ \emph {et~al.}(2008)\citenamefont {Houck},
  \citenamefont {Schreier}, \citenamefont {Johnson}, \citenamefont {Chow},
  \citenamefont {Koch}, \citenamefont {Gambetta}, \citenamefont {Schuster},
  \citenamefont {Frunzio}, \citenamefont {Devoret}, \citenamefont {Girvin},\
  and\ \citenamefont {Schoelkopf}}]{houck2008}%
  \BibitemOpen
  \bibfield  {author} {\bibinfo {author} {\bibfnamefont {A.~A.}\ \bibnamefont
  {Houck}}, \bibinfo {author} {\bibfnamefont {J.~A.}\ \bibnamefont {Schreier}},
  \bibinfo {author} {\bibfnamefont {B.~R.}\ \bibnamefont {Johnson}}, \bibinfo
  {author} {\bibfnamefont {J.~M.}\ \bibnamefont {Chow}}, \bibinfo {author}
  {\bibfnamefont {J.}~\bibnamefont {Koch}}, \bibinfo {author} {\bibfnamefont
  {J.~M.}\ \bibnamefont {Gambetta}}, \bibinfo {author} {\bibfnamefont {D.~I.}\
  \bibnamefont {Schuster}}, \bibinfo {author} {\bibfnamefont {L.}~\bibnamefont
  {Frunzio}}, \bibinfo {author} {\bibfnamefont {M.~H.}\ \bibnamefont
  {Devoret}}, \bibinfo {author} {\bibfnamefont {S.~M.}\ \bibnamefont
  {Girvin}},\ and\ \bibinfo {author} {\bibfnamefont {R.~J.}\ \bibnamefont
  {Schoelkopf}},\ }\href {https://doi.org/10.1103/physrevlett.101.080502}
  {\bibfield  {journal} {\bibinfo  {journal} {Physical Review Letters}\
  }\textbf {\bibinfo {volume} {101}},\ \bibinfo {pages} {080502} (\bibinfo
  {year} {2008})}\BibitemShut {NoStop}%
\bibitem [{\citenamefont {Yablonovitch}(1987)}]{yablonovitch1987}%
  \BibitemOpen
  \bibfield  {author} {\bibinfo {author} {\bibfnamefont {E.}~\bibnamefont
  {Yablonovitch}},\ }\href {https://doi.org/10.1103/physrevlett.58.2059}
  {\bibfield  {journal} {\bibinfo  {journal} {Physical Review Letters}\
  }\textbf {\bibinfo {volume} {58}},\ \bibinfo {pages} {2059} (\bibinfo {year}
  {1987})}\BibitemShut {NoStop}%
\bibitem [{\citenamefont {Painter}\ \emph {et~al.}(1999)\citenamefont
  {Painter}, \citenamefont {Lee}, \citenamefont {Scherer}, \citenamefont
  {Yariv}, \citenamefont {O{\textquotesingle}Brien}, \citenamefont {Dapkus},\
  and\ \citenamefont {Kim}}]{painter1999}%
  \BibitemOpen
  \bibfield  {author} {\bibinfo {author} {\bibfnamefont {O.}~\bibnamefont
  {Painter}}, \bibinfo {author} {\bibfnamefont {R.~K.}\ \bibnamefont {Lee}},
  \bibinfo {author} {\bibfnamefont {A.}~\bibnamefont {Scherer}}, \bibinfo
  {author} {\bibfnamefont {A.}~\bibnamefont {Yariv}}, \bibinfo {author}
  {\bibfnamefont {J.~D.}\ \bibnamefont {O{\textquotesingle}Brien}}, \bibinfo
  {author} {\bibfnamefont {P.~D.}\ \bibnamefont {Dapkus}},\ and\ \bibinfo
  {author} {\bibfnamefont {I.}~\bibnamefont {Kim}},\ }\href
  {https://doi.org/10.1126/science.284.5421.1819} {\bibfield  {journal}
  {\bibinfo  {journal} {Science}\ }\textbf {\bibinfo {volume} {284}},\ \bibinfo
  {pages} {1819} (\bibinfo {year} {1999})}\BibitemShut {NoStop}%
\bibitem [{\citenamefont {Zoepfl}\ \emph {et~al.}(2017)\citenamefont {Zoepfl},
  \citenamefont {Muppalla}, \citenamefont {Schneider}, \citenamefont
  {Kasemann}, \citenamefont {Partel},\ and\ \citenamefont
  {Kirchmair}}]{zoepfl2017characterization}%
  \BibitemOpen
  \bibfield  {author} {\bibinfo {author} {\bibfnamefont {D.}~\bibnamefont
  {Zoepfl}}, \bibinfo {author} {\bibfnamefont {P.~R.}\ \bibnamefont
  {Muppalla}}, \bibinfo {author} {\bibfnamefont {C.}~\bibnamefont {Schneider}},
  \bibinfo {author} {\bibfnamefont {S.}~\bibnamefont {Kasemann}}, \bibinfo
  {author} {\bibfnamefont {S.}~\bibnamefont {Partel}},\ and\ \bibinfo {author}
  {\bibfnamefont {G.}~\bibnamefont {Kirchmair}},\ }\href
  {https://aip.scitation.org/doi/full/10.1063/1.4992070} {\bibfield  {journal}
  {\bibinfo  {journal} {AIP Advances}\ }\textbf {\bibinfo {volume} {7}},\
  \bibinfo {pages} {085118} (\bibinfo {year} {2017})}\BibitemShut {NoStop}%
\bibitem [{\citenamefont {Dalmonte}\ \emph {et~al.}(2015)\citenamefont
  {Dalmonte}, \citenamefont {Mirzaei}, \citenamefont {Muppalla}, \citenamefont
  {Marcos}, \citenamefont {Zoller},\ and\ \citenamefont
  {Kirchmair}}]{dalmonte2015realizing}%
  \BibitemOpen
  \bibfield  {author} {\bibinfo {author} {\bibfnamefont {M.}~\bibnamefont
  {Dalmonte}}, \bibinfo {author} {\bibfnamefont {S.}~\bibnamefont {Mirzaei}},
  \bibinfo {author} {\bibfnamefont {P.}~\bibnamefont {Muppalla}}, \bibinfo
  {author} {\bibfnamefont {D.}~\bibnamefont {Marcos}}, \bibinfo {author}
  {\bibfnamefont {P.}~\bibnamefont {Zoller}},\ and\ \bibinfo {author}
  {\bibfnamefont {G.}~\bibnamefont {Kirchmair}},\ }\href
  {https://journals.aps.org/prb/abstract/10.1103/PhysRevB.92.174507} {\bibfield
   {journal} {\bibinfo  {journal} {Physical Review B}\ }\textbf {\bibinfo
  {volume} {92}},\ \bibinfo {pages} {174507} (\bibinfo {year}
  {2015})}\BibitemShut {NoStop}%
\end{thebibliography}%


\end{document}